\documentclass[lettersize,journal]{IEEEtran}

\usepackage[utf8]{inputenc} 
\usepackage[T1]{fontenc}
\usepackage{url}
\usepackage{ifthen}
\usepackage{cite}
\usepackage[cmex10]{amsmath} 
\usepackage{amsmath,amsfonts}
\usepackage{algorithm}
\usepackage{array}
\usepackage{textcomp}
\usepackage{stfloats}
\usepackage{url}
\usepackage{verbatim}
\usepackage{graphicx}
\usepackage{cite}
\usepackage{algorithmicx,algorithm}
\usepackage{algpseudocode}
\usepackage{bm}
\usepackage{verbatim}
\usepackage{booktabs}
\usepackage{amsmath,amssymb,amsfonts}
\usepackage{threeparttable}
\usepackage{multirow}
\usepackage{stackengine}

\usepackage[caption=false,font=normalsize,labelfont=sf,textfont=sf]{subfig}

\usepackage{hyperref}
\hypersetup{colorlinks,
	linkcolor=blue,%
	citecolor=blue}

\begin{document}



\title{Transformer-aided Wireless Image Transmission with Channel Feedback}
   


\author{
Haotian Wu,~\IEEEmembership{Graduate Student Member,~IEEE},
Yulin~Shao,~\IEEEmembership{Member,~IEEE},
Emre Ozfatura,~\IEEEmembership{Member,~IEEE},
Krystian Mikolajczyk,~\IEEEmembership{Senior Member,~IEEE},
Deniz~G\"und\"uz,~\IEEEmembership{Fellow,~IEEE}
\thanks{The authors are with the Department of Electrical and Electronic Engineering, Imperial College London, London SW7 2AZ, U.K. (e-mails: \{haotian.wu17, y.shao, m.ozfatura, k.mikolajczyk, d.gunduz\}@imperial.ac.uk).
}
\thanks{
This work received funding from the UKRI for the projects AI-R (ERC Consolidator Grant, EP/X030806/1) and SONATA (EPSRC-EP/W035960/1).
}
}

\maketitle

\begin{abstract}
This paper presents a novel wireless image transmission paradigm that can exploit feedback from the receiver, called  {JSCCformer-f}. We consider a block feedback channel model, where the transmitter receives noiseless/noisy channel output feedback after each block. The proposed scheme employs a single encoder to facilitate transmission over multiple blocks, refining the receiver's estimation at each block. Specifically, the unified encoder of JSCCformer-f can leverage the semantic information from the source image, and acquire channel state information and the decoder's current belief about the source image from the feedback signal to generate coded symbols at each block. Numerical experiments show that our JSCCformer-f scheme achieves state-of-the-art performance with robustness to noise in the feedback link. Additionally, JSCCformer-f can adapt to the channel condition directly through feedback without the need for separate channel estimation. We further extend the scope of the JSCCformer-f approach to include the broadcast channel, which enables the transmitter to generate broadcast codes in accordance with signal semantics and channel feedback from individual receivers.
\end{abstract}

\section{Introduction}
The evolution of wireless communication systems has empowered many new applications in Internet-of-things (IoT) and edge intelligence \cite{9261169,lo2023collaborative}, from autonomous vehicles to virtual reality, for which high information content signals (images, videos, LIDAR measurements) must be delivered with high quality and low latency. The conventional approach, which relies on Shannon's separation theorem, involves the independent design of source and channel coding, which is optimal in the asymptotic limit of infinite block-length \cite{shannon1948mathematical} for ergodic source and channel distributions. Although the separate design is known to be suboptimal in the practical finite block length regime, designing joint coding schemes is extremely challenging, and no such codes have been known that can perform on par with state-of-the-art separation-based codes.

In recent years, inspired by the tremendous success of deep learning (DL) techniques in many long-standing open problems, researchers have started to exploit deep neural networks (DNNs) to design novel and competitive joint source-channel coding (JSCC) schemes to surpass the performance of conventional separation-based schemes in practice \cite{bourtsoulatze2019deep,yoo2023role,9714510,kurka2020deepjscc,hu2023robust,DTAT,kurka2021bandwidth,9541059,10094735,9791398, tung2021deepwive}. Benefiting from the powerful task-oriented feature extraction capability of convolutional neural networks (CNNs), these DL-based JSCC schemes, commonly referred to as DeepJSCC, exhibit promising performance results, and even outperform the conventional separation-based schemes, particularly in the low signal-to-noise ratio (SNR) and low bandwidth regimes. The success of DeepJSCC schemes also extends to other downstream tasks beyond signal reconstructions, such as image retrieval \cite{9261169,wu2023features},  visual question answering~\cite{xie2022task}, edge inference \cite{shao2021learning}, or channel state information feedback~\cite{Mashhadi:ICASSP:20}, where the transmitter only transmits the relevant semantic features instead of all the source information, saving much bandwidth.

This paper investigates the JSCC problem in the presence of feedback signals from the receiver to the transmitter. In the asymptotic regime, the optimality of separation continues to hold in the presence of feedback\cite{cover1999elements}. Moreover, feedback does not increase the capacity of a memoryless channel\cite{shannon1956zero,dobrushin1958information}. However, feedback can simplify the coding mechanism, as demonstrated in\cite{schalkwijk1966coding,schalkwijk1967transmission}. In addition, in the practical finite blocklength regime, the optimal transmission scheme is wide open also in the presence of feedback. Interestingly, in \cite{schalkwijk1967transmission}, it was shown that when transmitting a single Gaussian source sample over finitely many uses of a Gaussian channel, a JSCC scheme with perfect channel output feedback is shown to achieve the optimal performance, although the optimal performance without feedback remains unknown. This shows that feedback in practice can be highly beneficial in both improving the performance and simplifying the coding scheme\cite{kostina2017joint}. 

It is noteworthy that only a limited number of studies have addressed the JSCC problem with channel feedback in the literature. In most of the earlier existing works, a scalable compression scheme is considered for the source (e.g., image or video), and an unequal error protection scheme is employed for the channel code. These schemes typically employ separate source and channel codes, with their rates adjusted based on channel feedback. For instance, \cite{chande1998joint} and \cite{755663} exploit one-bit acknowledgment/negative acknowledgment (ACK/NACK) feedback to adjust the channel code rate. In \cite{750824}, channel state information (CSI) is considered as feedback to enable the transmission of the source signal in a multicarrier scenario. In\cite{puri1998joint}, video signals are considered, and an automatic repeat request (ARQ) scheme is proposed to jointly allocate source and channel rates to video blocks. Similar rate allocation problems have also been studied in \cite{chou2000fec,chakareski2002computing,taubman2005optimal} for streaming media.

\begin{figure}[t] 
    \centering
    \includegraphics[scale=0.5]{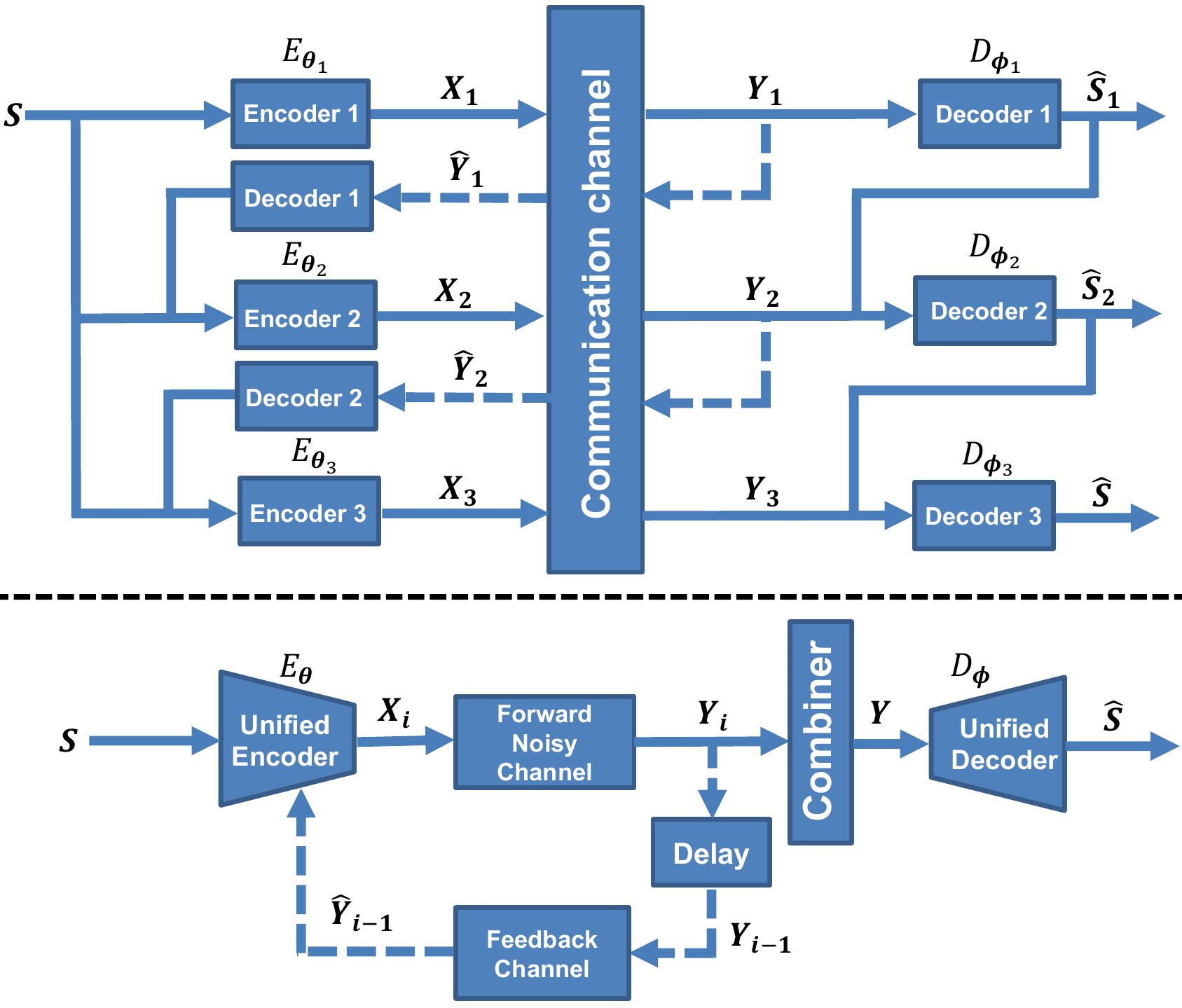}
    \caption{Illustration of alternative JSCC schemes for channels with feedback, where the solid lines represent forward channels, while the dotted lines represent feedback channels. 
     {Top: Illustration of the DeepJSCC-f scheme in \cite{kurka2020deepjscc} (with $m=3$ blocks as an example), where multiple independent encoders and decoders are trained as the channel codes in each block.} Bottom: Illustration of our proposed JSCCformer-f scheme, where a single unified encoder-decoder pair is employed at each block, significantly reducing the training and memory requirements as well as the coding complexity.}
    \label{fig_1}
\end{figure}

The first truly joint coding scheme for transmitting images over a wireless channel with feedback was proposed in \cite{kurka2020deepjscc}. The authors presented a stacked autoencoder structure, called DeepJSCC-f, to transmit the image in multiple blocks. As illustrated in Fig. \ref{fig_1}, DeepJSCC-f utilizes a set of encoders and decoders in the transmitter and receiver, respectively, and a different pair of encoder and decoder is used at each iteration. It is shown in\cite{kurka2020deepjscc} that feedback can improve the performance of CNN-based DeepJSCC, and the proposed DeepJSCC-f scheme achieves state-of-the-art performance. Even more significant gains are exhibited in \cite{kurka2020deepjscc} when variable-length coding is considered. 

The design of DeepJSCC-f, however, exhibits a few limitations that we wish to address in this paper.

\begin{itemize}
    \item \textbf{High complexity}. DeepJSCC-f trains and employs independent encoders and decoders for each interaction between the transmitter and receiver. This results in a substantial surge in both training complexity and memory requirements when the number of feedback blocks increases. A practical JSCC scheme should be able to effectively exploit channel feedback with low complexity.

    \item \textbf{Inadaptability}. Adaptability to varying channel conditions is another critical property for learning-based JSCC schemes. Specifically, training encoder and decoder network parameters for specific channel conditions, as demonstrated in prior works \cite{bourtsoulatze2019deep, kurka2020deepjscc, kurka2021bandwidth}, would necessitate the storage of multiple sets of network parameters. Alternatively, a JSCC scheme can achieve adaptability to channel conditions by leveraging the attention mechanism and training at random SNRs~\cite{wu2022vision,xu2021wireless, wu2022channel}. This approach enables dynamically scaling the features according to the channel conditions. Furthermore, it should be noted that all of these methods require the acquisition and feedback of CSI, which may not always be feasible. To the best of our knowledge, the direct learning of channel adaptability from noisy channel feedback has not yet been explored.

    \item \textbf{Suboptimality}. The training of DeepJSCC-f is conducted in a block-by-block fashion, where pairs of encoder and decoder are trained progressively. This leads not only to training instability, but also to suboptimal performance. In addition, the advent of novel DL architectures\cite{zheng2021rethinking,dosovitskiy2020image,chu2021twins} with respect to image processing and natural language processing, has rendered the CNN-enabled encoder and decoder as suboptimal alternatives.
    
    \item \textbf{Non-generalizability}. DeepJSCC-f was designed exclusively for point-to-point channels with feedback. On the other hand, we anticipate a more versatile encoding and decoding JSCC scheme that can be generalized to other communication scenarios, such as multiple access channels and broadcast channels, with feedback mechanisms.
\end{itemize}

This paper presents a novel JSCC paradigm with feedback, called JSCCformer-f, that effectively addresses the concerns outlined above. Unlike DeepJSCC-f, JSCCformer-f utilizes a unified encoder-decoder pair at the transmitter and receiver, respectively. Specifically, the transmission of source signals is divided into multiple blocks, and the channel feedback from previous blocks is used to facilitate the encoding process. The unified encoder leverages the semantic information from the source image, the CSI, and the feedback (i.e., the decoder's current belief about the source image) to generate coded symbols at each interaction. These coded symbols are then transmitted to the receiver to refine its belief and mitigate the impact of channel noise in subsequent interactions. Finally, the unified decoder accumulates the received successive coded blocks and reconstructs the source image after the last round of interaction. Overall, an image is transmitted to the receiver with multiple blocks in a coarse-to-fine fashion, and the well-trained model exhibits channel adaptability, thanks to the encoder's ability to attend to feedback and CSI.

Furthermore, JSCCformer-f is built upon the vision transformer (ViT) architecture\cite{dosovitskiy2020image,zheng2021rethinking} as the unified encoder and decoder. Unlike CNNs that learn the semantic features of images from a local to global context with a limited receptive field, ViTs employ a global self-attention (SA) mechanism for a more discriminative representation of semantic features. This transformer-based SA strategy naturally fits our encoding and decoding process with feedback, where we rethink the image-to-channel mapping with interactions from a sequences-to-sequence perspective, inspired by the feedback channel code design in \cite{GBAFC}.

Our main contributions can be summarized as follows:
\begin{itemize}
\item We present a new feedback-aided JSCC architecture, JSCCformer-f, for wireless image transmission with channel feedback. JSCCformer-f exploits the semantics of the source image and the available feedback signal in each interaction to generate coded symbols and to refine the decoder's belief about the source image. JSCCformer-f introduces a new way of utilizing the semantics and the feedback through an SA mechanism of the transformer architecture. Although it is used here for image transmission, this framework is general and can be used to convey any desired semantic information\cite{gunduz2022beyond,9530497}

\item Our scheme is a resource-friendly pipeline, employing only a single unified encoder to accommodate a wide range of interaction numbers without introducing additional parameters or training effort. Flexible strategies, such as variable rate transmission, are also enabled to allow a considerable economy of resources for attaining a specific quality objective of the transmission. The proposed JSCCformer-f scheme can also directly learn to adapt the various channel conditions from the noiseless/noisy channel feedback and achieves considerable gains compared to all the previous methods.

\item Numerical experiments verify that our JSCCformer-f scheme achieves state-of-the-art in all SNRs and bandwidth scenarios. We also show its robustness to feedback channel noise. 

\item To demonstrate the generalizability of the proposed scheme, we extend the JSCCformer-f framework to a broadcast channel with two receivers, where the encoder at the transmitter has to attend to the feedback from both receivers and generate coded symbols accordingly.
\end{itemize}

\section{Problem Formulation}
We consider the problem of wireless transmission of images in the presence of noisy channel output feedback. The goal is to transmit an input image $\bm{S}\in \mathbb{R}^{h\times w\times 3}$ over a noisy communication channel with the help of the channel output feedback, where $h,w$ and $3$ denote the height, width, and color channels of an RGB image. The transmission of each image is done in $m$ blocks, where $m=1$ corresponds to the no-feedback scenario. We assume that after each block, the channel output is fed back to the transmitter via a noisy or noiseless feedback link. At each block, the transmitter can exploit the feedback signals it has received up until that block.

Accordingly,  {for a fixed total bandwidth allocation,} a JSCC code that transmits an image of $n$ pixel intensities in $m$ blocks, each consisting of $k$ channel symbols, is denoted by the triplet $(n,m,k)$. Here $n=3hw$ denotes the source bandwidth, while $mk$ denotes the total channel bandwidth  {cost. We note that, for a fixed bandwidth ratio $R$, an increase in the number of blocks leads to a reduction in the number of channel uses per block.} Following\cite{kurka2020deepjscc}, we define the \textit{bandwidth ratio} as $R=mk/n$, representing the average number of available channel symbols per source dimension.

As shown in Fig \ref{fig_1}, in the $i$-th transmission block, the encoder function ${E}_{\bm{\theta_{i}}}$, parameterized by $\bm{\theta_i}$, maps $\bm{S}$ and all the available feedback signals into a complex channel vector $\bm{X_i}=E_{\bm{\theta}_i}(\bm{S},\bm{\hat{Y}_{1:i-1})}\in \mathbb{C}^{k}$, where $\bm{\hat{Y}_{1:i-1}}\triangleq [\bm{\hat{Y}_{1}},{\ldots},\bm{\hat{Y}_{i-1}}]$ denotes all the previous feedback signals available at the $i$-th transmission block, and $\bm{\hat{Y}_{i}}\in \mathbb{C}^{k}$ is the channel output feedback received after the $i$-th block. Note that the encoder, in general, can consist of $m$ encoding functions, one for each block. The transmitted signal $\bm{X_i}$ is subject to a power constraint $P_s$ as: 
\begin{equation}
\frac{1}{mk}\sum_{i=1}^m \mathbb{E}\left[\|\bm{X_i}\|^2_\text{2}\right]\leq P_s,
\end{equation} where we set $P_s=1$ without loss of generality.


The forward channel in the $i$-th transmission block is modeled as an additive white Gaussian noise (AWGN) channel with fading as:
\begin{equation}
   \bm{{Y}_{i}}=\mathcal{H}(\bm{X_i})= \begin{cases}
h\bm{X_i}+\bm{W_i}, & \text{slow fading channel}\\
\bm{X_i}+\bm{W_i}, & \text{AWGN channel},\\
\end{cases}
\end{equation}
where $\bm{{Y}_{i}}\in \mathbb{C}^{k}$ is the channel output in $i$-th forward transmission block, $h\in \mathbb{C}$ is the channel gain, and $\bm{W_i}\in \mathbb{C}^{k}$ is the AWGN term. Each element of $\bm{W_i}$ is sampled from independent and identically distributed (i.i.d.) complex Gaussian noise as $W[i] \sim \mathcal{CN}(0,\sigma_w^2)$. We consider a slow fading scenario and assume that $h$ remains constant for the whole duration ($mk$ channel symbols) of the transmission of one image, but takes an independent realization for each image. We assume that $h$ is sampled from an i.i.d. complex Gaussian distribution, i.e., $h\sim \mathcal{CN}(0,\sigma_h^2)$. For the static AWGN scenario, we set $h=1$. In order to measure the channel quality, {we define the SNR as $10\log_{10} \frac{\sigma_h^2}{\sigma_w^2} (\text{dB})$}.


After each block, the channel output vector $\bm{Y_i}$ is then fed back to the transmitter through a feedback link, which can be either perfect or noisy, modeled as:
\begin{equation}
   \bm{\hat{Y}_{i}}=\mathcal{H}_f(\bm{Y_{i}})= \begin{cases}
\bm{Y_{i}}, & \text{perfect feedback link}\\
\bm{Y_{i}}+\bm{W_f}, & \text{AWGN feedback link},\\
\end{cases}
\end{equation}
where $\bm{W_f}\in \mathbb{C}^{k}$ is the AWGN term sampled from an i.i.d. additive complex Gaussian distribution, i.e., $W_f[i] \sim \mathcal{CN}(0,\sigma_f^2)$. 

The decoding function $D_{\bm{\phi}}$, parameterized by $\bm{\phi}$, reconstructs the transmitted image as $\bm{\hat{S}}=D_{\bm{\phi}}(\bm{Y})\in \mathbb{R}^{h\times w\times 3}$, where $\bm{Y}\triangleq[\bm{Y_1},{\ldots},\bm{Y_m}]$.

 {The reconstruction quality is quantified through the peak signal-to-noise ratio (PSNR) and the learned perceptual image patch similarity (LPIPS) metrics\cite{8578166}. The PSNR metric serves as an indicator of image distortion at the per-pixel level. A higher PSNR value is indicative of a better image reconstruction quality}, which is given as:
\begin{equation}
    \text{PSNR} (\bm{S},\bm{\hat{S}})=10\log_{10}\frac{(\max{\bm{S}})^2}{\text{MSE}(\bm{S},\bm{\hat{S}})}~(\text{dB}),
\end{equation}
where $\max {\bm{S}}$ denotes the maximum possible value of the input signal $\bm{S}$. The mean squared error (MSE) between $\bm{S}$ and its estimate $\bm{\hat{S}}$ is defined as $\text{MSE}(\bm{S},\bm{\hat{S}})\triangleq \frac{1}{3hw}\|\bm{S}-\bm{\hat{S}}\|^2_2$.  {LPIPS computes the dissimilarity within the feature space between the input image and the reconstruction, employing a deep neural network, such as VGG. LPIPS exhibits a superior alignment with human perception compared to conventional pixel-wise metrics\cite{8578166}. A lower LPIPS score implies a better perceptual quality, indicative of a greater degree of perceptual similarity observed among image patches, defined as:}
 {
\begin{equation}
    \text{LPIPS} (\bm{S},\bm{\hat{S}})=\sum_L \frac{1}{H_lW_l}\sum_{h,w}\|\bm{w_l}\odot(\bm{{y}^l}-\bm{\hat{y}^l})\|^2_2,
\end{equation}
where $\bm{{y}^l}, \bm{\hat{y}^l}\in \mathbb{R}^{H_l\times W_l\times C_l}$ are the intermediate features derived from the $l$-th layer of the given network (VGG) composed of a total of $L$ layers. $H_l, W_l,$ and $C_l$ are the intermediate feature map's height, width, and channel dimension. $\bm{w^l}\in \mathbb{R}^{C_l}$ is the weight vector, and $\odot$ is the channel-wise feature multiplication operation\cite{8578166}.}

Given the code parameters $(n,m,k)$, we aim to optimize the encoder and decoder parameters $\bm{\theta_1}, \dots, \bm{\theta_m}$ and $\bm{\phi}$, respectively, with the objective of improving the average reconstruction performance at the receiver.

\section{JSCCformer-f Architecture}
\begin{figure*}[t] 
    \centering
    \includegraphics[scale=0.7]{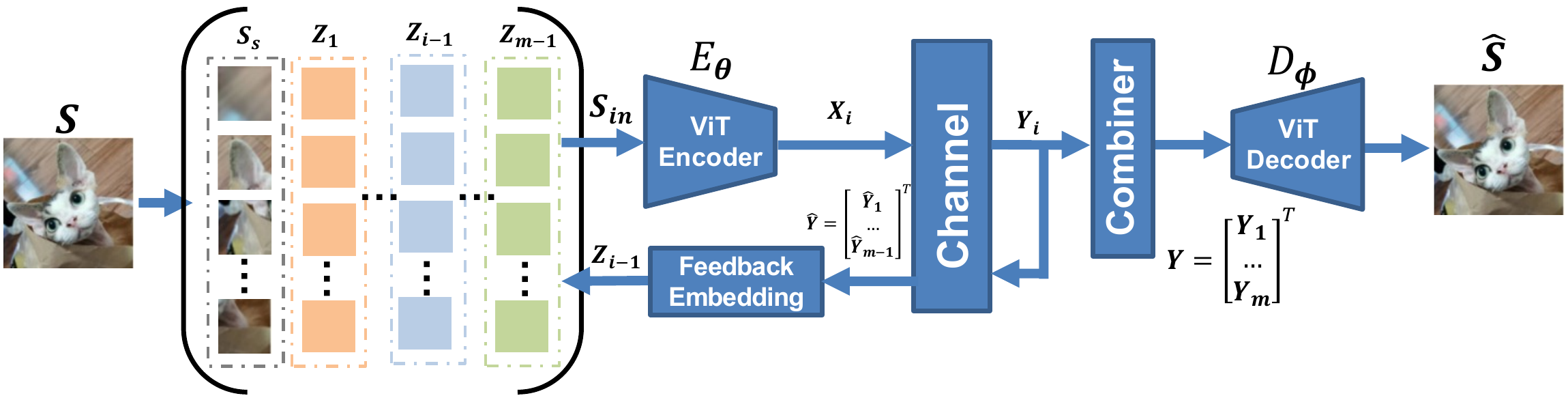}
    \caption{ {The pipeline of our JSCCformer-f scheme. In the $i$-th transmission block, the ViT-based encoder $E_{\bm{\theta}}$ encodes the channel symbols $\bm{X_i}$  {based on the input image and the channel feedback signals received until that time, where $\bm{\hat{Y}_{{i-1}}}$ is the noisy channel feedback for the $i-1$-th transmission and $\bm{Z_{i-1}}$ is its embedding. Note that $(\bm{\hat{Y}_{{i}}}, \ldots, \bm{\hat{Y}_{{m}}})$ and $(\bm{Z_{i}}, \ldots, \bm{Z_{m-1}})$ are padded with zeros as they correspond to future channel feedback signals.}}}
    \label{Secom_pipeline}
\end{figure*}

This section introduces the proposed feedback-aided JSCC paradigm for an image transmission task, denoted by JSCCformer-f, depicted in Fig \ref{Secom_pipeline}. Transmission is partitioned into $m$ blocks, each corresponding to one interaction between the transmitter and receiver.
JSCCformer-f utilizes a pair of ViTs, which possess symmetric inner structures, as the encoder and decoder, respectively. The architecture of JSCCformer-f is illustrated in Figure \ref{encoder}, and consists of three key components: ViT encoding, ViT decoding, and the loss function.

\subsection{{ViT-encoder}}
The encoding process entails sequentializing the image, embedding the feedback signal, and utilizing a ViT-based encoder $E_{\bm{\theta}}$ to map the updated input sequence to channel symbols. The encoding process is elaborated in four steps: (1) image-to-sequence transformation, (2) feedback embedding, (3) input sequence update, and (4) ViT-based encoding. The encoding process is presented in Algorithm \ref{encoder_algorithm} as a pseudocode.

\subsubsection{Image-to-sequence transformation}
To construct the input of JSCCformer-f at each block, we first convert the three-dimensional input image $\bm{S}$ into a sequence of vectors, denoted by $\bm{S_s}$. Specifically, given a source image $\bm{S}\in\mathbb{R}^{h\times w\times 3}$, we divide $\bm{S}$ into a grid of $p\times p$ patches, and flatten the pixel intensities of each patch to form a sequence of vectors of dimension $\mathbb{R}^{\frac{3hw}{p^2}}$. In this way, $\bm{S}$ is converted to $\bm{S_s}\in\mathbb{R}^{l\times c}$, where $l={p^2}$ is the sequence length and $c\triangleq\frac{3hw}{p^2}$ is the dimension of each vector.  {Here $p$ serves as a hyperparameter, where a higher value of $p$ indicates a more detailed partition of the image into smaller patches, facilitating more fine-grained feature learning but at the expense of a corresponding rise in computational complexity\cite{wu2023transformer}.}

\subsubsection{{Feedback embedding}} Within each transmission block, a feedback embedding module is employed at the transmitter, parameterized as $f_e(\cdot): \mathbb{R}^{l\times \frac{2mk}{l}}\rightarrow \mathbb{R}^{l\times z}$, to embed all the received feedback signals $\bm{\hat{Y}}$ into $\bm{Z_i}$ for the subsequent ViT-encoding step, where $z$ is the output dimension for the feedback embedding block.

We propose two different feedback embedding schemes, JSCCformer-f and JSCCformer-f (lite), respectively.
Specifically,  {the embedding scheme of JSCCformer-f} utilizes an additional ViT-based decoder  {at the transmitter} (the weights of which are copied from the decoder at the receiver) to estimate the receiver's current belief from the received feedback, which is then combined with the raw received feedback as the input to the encoder.  {During the training phase, the gradient over the decoder is stopped at the transmitter to simplify the training process, especially in the application of variable rate and successive refinement transmission.} The second scheme, in contrast, directly feeds the received channel feedback to the encoder. That is why JSCCformer-f with the second embedding scheme is called JSCCformer-f (lite).

For JSCCformer-f and JSCCformer-f (lite), we have $z=\frac{2k}{l}+c$ and $z=\frac{2k}{l}$, respectively.
Thanks to the integration of receiver's belief, JSCCformer-f achieves a better performance at the cost of additional complexity.

The detailed formula of $\bm{Z_i}=f_e(\bm{\hat{Y}})$ is provided below:
 {
\begin{equation}
   \bm{Z_i}= \begin{cases}
\textit{concat}(\textit{sg}(D_{\bm{\phi}}(\bm{\hat{Y}})),\bm{\hat{Y}_{i}}), & \text{JSCCformer-f},\\
\bm{\hat{Y}_{i}}, & \text{JSCCformer-f (lite)},
\end{cases}
\label{fb_emb}
\end{equation}}where $\bm{\hat{Y}_{i}}\in \mathbb{R}^{l\times \frac{2k}{l}}$ is the channel output feedback of the $i$-th block, reshaped from $\mathbb{C}^{k}$, $\bm{\hat{Y}}=\bm{\hat{Y}_{1:m}}\triangleq [\bm{\hat{Y}_1}, \ldots, \bm{\hat{Y}_{m}}]\in \mathbb{R}^{l\times \frac{2mk}{l}}$,  {$\textit{sg}(\cdot)$ means stop gradient operation}, $\textit{concat}(\cdot)$ denotes concatenation, and $\bm{Z_i}\in \mathbb{R}^{l\times z}$ is the embedded feedback matrix given $\bm{\hat{Y}_{i}}$ for the $(i+1)\text{th}$ block. We pad $\bm{\hat{Y}_t}$ with a matrix of zeros when there is no knowledge of $\bm{\hat{Y}_t}$. That is, in the $i$-th block of a total $m$ block transmission, $\bm{\hat{Y}_{1:i-2}}$ is given, $\bm{\hat{Y}_{i-1}}$ is updated, while $\bm{\hat{Y}_{i:m}}$ is set to zeros.

\subsubsection{{Input sequence update}}
As shown in Algorithm \ref{encoder_algorithm}, we concatenate $\bm{S_s}$ and the embedded feedback matrix $\bm{Z_{1:m-1}}\triangleq [\bm{Z_{1}},\ldots,\bm{Z_{m-1}}]$ to form the input sequence $\bm{S_{in}}$ of the ViT-based encoder in the $i$-th block as: 
\begin{equation}
    \bm{S_{in}}=\textit{concat}(\bm{S_s},\bm{Z_{1:m-1}}),
\end{equation}
 where $\bm{Z_{1:m-1}}\in \mathbb{R}^{l\times (m-1)z}$ and $\bm{S_{in}}\in \mathbb{R}^{l\times (c+(m-1)z)}$. 

In each transmission block ($i$-th block), we update $\bm{Z_{1:m-1}}$ and $\bm{S_{in}}$, once $\bm{Z_{i-1}}$ generated from Eqn. \eqref{fb_emb} with a new $\bm{\hat{Y}_{i-1}}$ becomes available. Similarly, if there is no knowledge of feedback $\bm{\hat{Y}_{t}}$ in the current $i$-th transmission block (i.e., $t\geq i$), $\bm{Z_k}$ would be padded with zeros. In particular, for the $i$-th block of a total $m$ transmission blocks, $\bm{Z_{i-1}}$ is updated while $\bm{Z_{i:m-1}}$ is set to zeros. 


\begin{figure*}[tb] 
    \centering
    \includegraphics[scale=0.68]{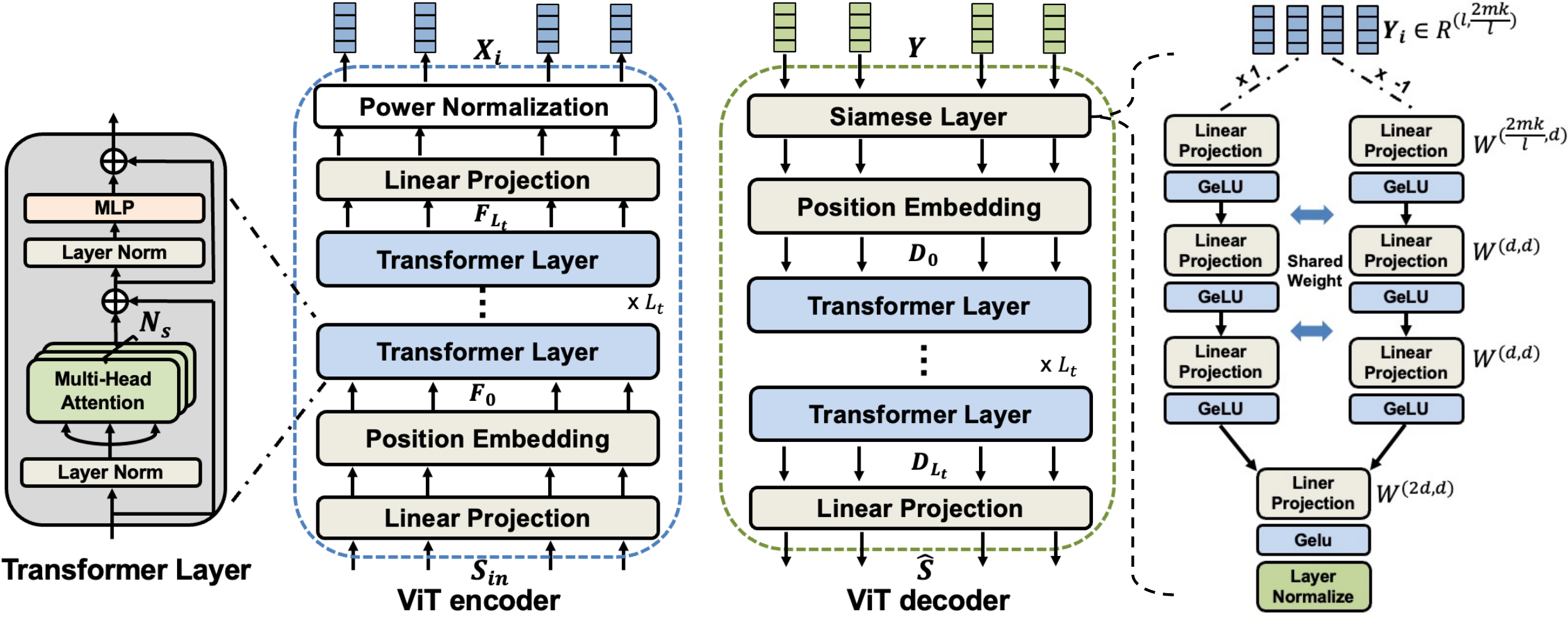}
    \caption{ {The architecture of the encoder and decoder, where a symmetric structure is designed to encode the input sequence and reconstruct the source signal.}}
    \label{encoder}
\end{figure*}

\subsubsection{{ViT-based encoding}} In each block, we use a ViT-based encoder $E_{\bm{\theta}}: \mathbb{R}^{{l\times (c+(m-1)z)}}\rightarrow \mathbb{C}^{k}$, parameterized by $\bm{\theta}$, to generate the channel symbols based on the updated $\bm{S_{in}}$. The architecture of our ViT-based encoder is shown in Fig. \ref{encoder}, mainly consisting of linear projection layers, a position embedding layer, several transformer layers, and a power normalization layer. A triplet $(L_t, N_s, d)$ can be utilized to characterize our proposed ViT-based encoder, wherein $L_t$, $N_s$, and $d$ represent the number of transformer layers, SA heads, and hidden units, respectively.

\textbf{Linear projection:}
The input matrix $\bm{S_{in}}$ first undergoes a linear projection with parameters $\bm{W_{0}}\in \mathbb{R}^{(c+(m-1)z)\times d}$, followed by a position embedding operation $P_{e}(\cdot): \mathbb{R}^{l\times d}\rightarrow \mathbb{R}^{l\times d}$, resulting in the initial input $\bm{F_{0}}\in \mathbb{R}^{l\times d}$:
\begin{equation}
\bm{F_{0}}=\bm{S_{in}}\bm{W_{0}}+\bm{p_{e}},
\label{pos_embed}
\end{equation}
where $d$ is the output dimension of the hidden projection layer, $\bm{F_{0}}\in \mathbb{R}^{l\times d}$ is the output of the linear projection and position embedding layer for the following transformer layers, and $\bm{p_{e}}\in \mathbb{R}^{l\times d}$ is the output of the position embedding (PE) operation $P_{e}(\cdot)$.
\begin{algorithm}[t] 
    \caption{ViT-encoding algorithm}
    \textbf{Input:} $\bm{S_s}\in \mathbb{R}^{l\times c}$\\
    \textbf{Output:} $\bm{X_{i}}\in \mathbb{R}^{l\times \frac{2k}{l}}, \bm{Y_{i}}\in \mathbb{R}^{l\times \frac{2k}{l}}$\\
    \textbf{For the $1$-st block} \\
    \vspace{-15pt}
    \begin{algorithmic}[1]
    \For{i= 1:m-1}
         \State \texttt{$\bm{\hat{Y}_{i}}=\bm{0_{l,\frac{2k}{l}}} \in \mathbb{R}^{l\times \frac{2k}{l}}$}
     \EndFor
     
     \State \texttt{$\bm{\hat{Y}}= [\bm{\hat{Y}_1}, \ldots, \bm{\hat{Y}_{m-1}}]\in \mathbb{R}^{l\times \frac{2k(m-1)}{l}}$
    }     
    \Comment{Initialize $\bm{\hat{Y}}$}
    \For{i= 1:m-1}
         \State \texttt{$\bm{Z_{i}}=\bm{0_{l,z}} \in \mathbb{R}^{l\times z}$}
     \EndFor
     \Comment{Initialize each $\bm{Z}_i$}
    \State \texttt{$\bm{Z_{1:m-1}}\triangleq [\bm{Z_{1}},\ldots,\bm{Z_{m-1}}]\in \mathbb{R}^{l\times (m-1)z}$}
    \State \texttt{$\bm{S_{in}}=concat(\bm{S_s},\bm{Z_{1:m-1}}) \in \mathbb{R}^{l\times c_{in}}$}
    \State \texttt{$\bm{X_1}=\bm{E_{\theta}}(\bm{S_{in}}) \in \mathbb{R}^{l\times \frac{2k}{l}}\rightarrow \mathbb{C}^{k}$}
    \State \texttt{$\bm{Y_1}=\mathcal{H}(\bm{X_1})\in \mathbb{C}^{k}\rightarrow \mathbb{R}^{l\times \frac{2k}{l}}$} 
    \Comment{Forward link}
    \end{algorithmic}
    \textbf{For the $i$-th block, where $i \in [2,m-1]$} \\
    \vspace{-15pt}
    \begin{algorithmic}[1]
    \State \texttt{$\bm{\hat{Y}_{i-1}}=\mathcal{H}_f(\bm{Y_{i-1}})\in \mathbb{C}^{k}\rightarrow \mathbb{R}^{l\times \frac{2k}{l}}$}     
    \Comment{Feedback link}
    \State \texttt{$\bm{\hat{Y}}= [\bm{\hat{Y}_1} \ldots \bm{\hat{Y}_{m-1}}]$
    }     
    \Comment{Update $\bm{\hat{Y}}$}
    \State \texttt{$\bm{Z_{i-1}}=f_{e}(\bm{\hat{Y}})$}
    \Comment{Update $\bm{Z_{i-1}}$}
    \State \texttt{$\bm{Z_{1:m-1}}= [\bm{Z_{1}},\ldots,\bm{Z_{m-1}}]$}
    \Comment{Update $\bm{Z_{1:m-1}}$}
    \State \texttt{$\bm{S_{in}}=concat(\bm{S_s},\bm{Z_{1:m-1}})$}
    \Comment{Update input $\bm{S}_{in}$}
    \State \texttt{$\bm{X_i}=\bm{E_{\theta}}(\bm{S_{in}}) \in \mathbb{R}^{l\times \frac{2k}{l}}\rightarrow \mathbb{C}^{k}$}
    \State \texttt{$\bm{Y_i}=\mathcal{H}(\bm{X_i})\in \mathbb{C}^{k}\rightarrow \mathbb{R}^{l\times \frac{2k}{l}}$} 
    \end{algorithmic}
    \label{encoder_algorithm}
    \end{algorithm}
    
{\textbf{Position embedding:}}
Two methods are adopted for the positional embedding $P_{e}(\cdot)$ of JSCCformer-f: a dense layer-based position embedding (DPE) \cite{zheng2021rethinking} and a conditional position embedding (CPE) \cite{chu2021twins}. 

DPE employs a dense layer to embed the index vector $\bm{p}$ of each patch into a $d$-dimensional vector, denoted as: $\bm{p_e}=P_e(\bm{p})$. On the other hand, CPE employs a 2-D convolution with a kernel size of $k=3$ and $\frac{k-1}{2}$ zero padding to generate positional encodings dynamically, conditioned on the local neighborhood of the input. To elaborate further, we first reshape the encoded feature $\bm{S_{in}W_0}\in \mathbb{R}^{l\times d}$ into a $2$-D image of dimension $\mathbb{R}^{p\times p\times d}$, which is then fed into a 2-dimensional convolution layer. The resultant outputs are then reshaped as $\bm{p_{e}}\in \mathbb{R}^{l\times d}$. The final result of the conditional position embedding can be written as: $\bm{p_e}=P_e(\bm{S_{in}W_0})$.

DPE is the default positional embedding scheme for JSCCformer-f. It is lightweight and yields satisfactory performance. In contrast, CPE requires more computations. 
It is worth pointing out that CPE is more suitable for high-resolution images\cite{wu2023transformer}. 

\textbf{Transformer Layer:}
As shown in Fig. \ref{encoder}, the intermediate feature map $\bm{F_{i}}$ is generated by the $i\textit{-}th$ transformer layer by a multi-head self-attention (MSA) block and a multi-layer perceptron (MLP) layer with a residual skip as:
\begin{equation}
\bm{F_{i}}=MSA(\bm{F_{i-1}})+MLP(MSA(\bm{F_{i-1}})),
\label{trans_layer_eqn}
\end{equation}
where $\bm{F_{i}}\in \mathbb{R}^{l\times d}$ is the output sequence of the $i$-th transformer layer, GeLU activation and layer normalization operations are applied before each MSA and MLP block. 

Each MSA block consists of $N_{s}$ SA modules with a residual skip, which can be formulated as:

\begin{equation}
MSA(\bm{F_{i}})=\bm{F_{i}}+[SA_1(\bm{F_{i}}),\ldots,SA_{N_{s}}(\bm{F_{i}})]\bm{W_i},
\label{msa}
\end{equation}
where the output of all the SA modules $SA(\bm{F_{i}})\in \mathbb{R}^{l\times d_{s}}$ are concatenated for a linear projection $\bm{W_i}\in \mathbb{R}^{d_{s}N_{s}\times d}$, and $d_s=d/N_{s}$ is the output dimension of each SA operation.

For each SA module, the operations can be formulated as: 
\begin{equation}
SA(\bm{F_{l-1}})=softmax(\frac{\bm{qk}^{T}}{\sqrt{d}})\bm{v},    
\end{equation}
where $\bm{q},\bm{k},\bm{v}\in \mathbb{R}^{l\times d_s}$ are the query, key, and value vectors, respectively,  generated from three linear projection layers $\bm{W_q, W_k, W_v}\in \mathbb{R}^{d\times d_s}$ as: 
\begin{equation}
   \bm{q}=\bm{F_{l-1}W_q},~~\bm{k}=\bm{F_{l-1}W_k},~~\bm{v}=\bm{F_{l-1}W_v}. 
\end{equation}

\textbf{Linear projection and power normalization:}
After $L_t$ transformer layers, we apply a linear projection $\bm{W_c}\in \mathbb{R}^{d\times \frac{2k}{l}}$ to map the output of the transformer layers $\bm{F_{L_t}}$ to channel symbols as: $\bm{\bar{X}_i}=\bm{F_{L_t}}\bm{W_c}$, 
where $\bm{\bar{X}_i}\in \mathbb{R}^{l\times \frac{2k}{l}}$ is then reshaped and normalized to satisfy the power constraint to form the channel input symbols $\bm{X_i}\in \mathbb{R}^{l\times \frac{2k}{l}}$.

\subsection{ViT decoder}
At the decoder, we first employ a combiner to form the input sequence for the ViT-decoder by concatenating all the received signals as: 
\begin{equation}
    \bm{{Y}}\triangleq [\bm{{Y}_1}, \ldots, \bm{{Y}_{m}}]\in \mathbb{R}^{l\times \frac{2mk}{l}}.
\end{equation}
While we use the concatenation operation as a combiner for simplicity, we could also use a learning-based method, such as a CNN, or a fully connected layer. 

Given $\bm{{Y}}$, a symmetrical ViT-based decoder $D_{\bm{\phi}}: \mathbb{R}^{{l\times 2mk}}\rightarrow \mathbb{R}^{l\times c}$, parameterized by $\bm{\phi}$ and the same triplet $(L_t, N_s, d)$ with the ViT-encoder, is designed to reconstruct the source image as $\bm{\hat{S}}=D_{\phi}(\bm{Y})$, which consists of a Siamese layer, position embedding layer, transformer layer and linear projection layer. Each component of the ViT decoder is explained next, also presented as a pseudocode in Algorithm \ref{decoder_algorithm}.

\textbf{Siamese layer and position embedding:} We employ a weight-shared Siamese layer $\text{Siam}(\cdot)$ consisting of several linear projection layers and GeLU activation functions. As shown in Fig. \ref{encoder}, $\bm{Y}$ multiplied by $1$ and $-1$ are fed into several linear projection layers and GeLU functions. We expect these GeLU functions and linear projection layers can learn to truncate excessive noise realizations  {to bootstrap the performance\cite{wu2023transformer}, similarly to the designs in \cite{shao2022attentioncode, GBAFC}}.

To introduce the position information to the decoding process, we add the same position embedding layer as in \eqref{pos_embed} to get the output $\bm{D_{0}}\in \mathbb{R}^{l\times d}$ as : 
\begin{equation}
    \bm{D_{0}}=\text{Siam}(\bm{S_d})+\bm{p_e},
\end{equation}
where $\text{Siam}(\cdot)$ denotes the operations of the Siamese layer, with parameters illustrated in Fig. \ref{encoder}, and $\bm{p_{e}}\in \mathbb{R}^{l\times d}$ is the output of the position embedding operation.
\begin{algorithm}[t] 
    \caption{ViT-based decoder algorithm}
    \textbf{For the transmitter of $i$-th block transmission:} \\
        \textbf{Input:} $\bm{\hat{Y}_{i-1}}\in \mathbb{R}^{l\times \frac{2k}{l}}$, \textbf{Output:} $\bm{\hat{S}_{si}}\in \mathbb{R}^{l\times c}$\\
    \vspace{-15pt}
    \begin{algorithmic}[1] 
     \State \texttt{$\bm{\hat{Y}}=[\bm{\hat{Y}_{1}},\ldots,\bm{\hat{Y}_{m-1}}]$}
      \Comment{Update $\bm{\hat{Y}}$ with $\bm{\hat{Y}_{i-1}}$}
        \State \texttt{$\bm{\hat{S}_{si}}=D_{\bm{\phi}}(\bm{\hat{Y}})\in \mathbb{R}^{l\times c}$}     \Comment{Reconstruct $\bm{S_s}$ for refining}
    \end{algorithmic}
    \vspace{5pt}
    \textbf{For the receiver:} \\
    \textbf{Input:} $\bm{{Y_i}}\in \mathbb{R}^{l\times \frac{2k}{l}}$, \textbf{Output:} $\bm{\hat{S}}\in \mathbb{R}^{h\times w\times 3}$\\
    \vspace{-15pt}
    \begin{algorithmic}[1]
    \For{i= 1:m}
        \If{i==1}{ 
        \texttt{$\bm{{Y}}=\bm{{Y}_1}$}\;
        }
        \Else {\texttt{ $\bm{{Y}}=concat(\bm{{Y}},\bm{{Y}_i})$}
        }
        \EndIf
     \EndFor
     \Comment{Combine all received $\bm{{Y}_i}$}
     \State \texttt{$\bm{\hat{S}}=D_{\phi}(\bm{Y})\in \mathbb{R}^{l\times c}\Rightarrow\mathbb{R}^{h\times w\times 3}$}     
     \Comment{Reconstruct the $\bm{S}$}
    \end{algorithmic}
    \label{decoder_algorithm}
    \end{algorithm}
    
\textbf{Transformer layer:}
After the Siamese layer and positional embedding, $\bm{D_{0}}$ is passed through $L_t$ transformer layers:
\begin{equation}
\bm{D_{l}}=MSA(\bm{D_{l-1}})+MLP(MSA(\bm{D_{l-1}})),
\end{equation} where $\bm{D_{i}}\in \mathbb{R}^{l\times d}$ is the output of the $i$-th transformer layer at the decoder, and the $MSA$ and $MLP$ blocks share the same structure as those in \eqref{trans_layer_eqn}.

\textbf{Linear Projection:}
Given the output of the $L_t$-transformer layer $\bm{D_{L_t}}$, we apply a linear projection $\bm{W_{out}}\in \mathbb{R}^{d\times c}$, and then reshape the output into a matrix of size $\mathbb{R}^{h\times w \times 3}$ to reconstruct the input image as: 
\begin{equation}
    \bm{\hat{S}}=\text{reshape}(\bm{D_{L_t}}\bm{W_{out}}).
\end{equation}

\subsection{{Loss function}}
The encoder and decoder are optimized jointly by minimizing the mean square error between the input image $\bm{S}$ and its reconstruction $\bm{\hat{S}}$ as: 
\begin{equation}
\mathcal{L}(\bm{\theta},\bm{\phi} )= \mathbb{E}\big[\text{MSE}(\bm{x},\bm{\hat{x}})\big] \triangleq \mathbb{E}\big[ \|\bm{S}-\bm{\hat{S}}\|^2_2\big],
\end{equation}
where the expectation is taken over the randomness both in the source pixels and the channel state. The aim is to find the parameters $(\bm{\theta}^*,\bm{\phi}^* )$ that minimize the loss function $\mathcal{L}(\bm{\theta},\bm{\phi})$.

\begin{figure*}[tb]
    \centering
    \subfloat[$R=1/6$]{
    \label{ViT_8} 
    \begin{minipage}[t]{0.32\linewidth}
    \centering
    \includegraphics[scale=0.3]{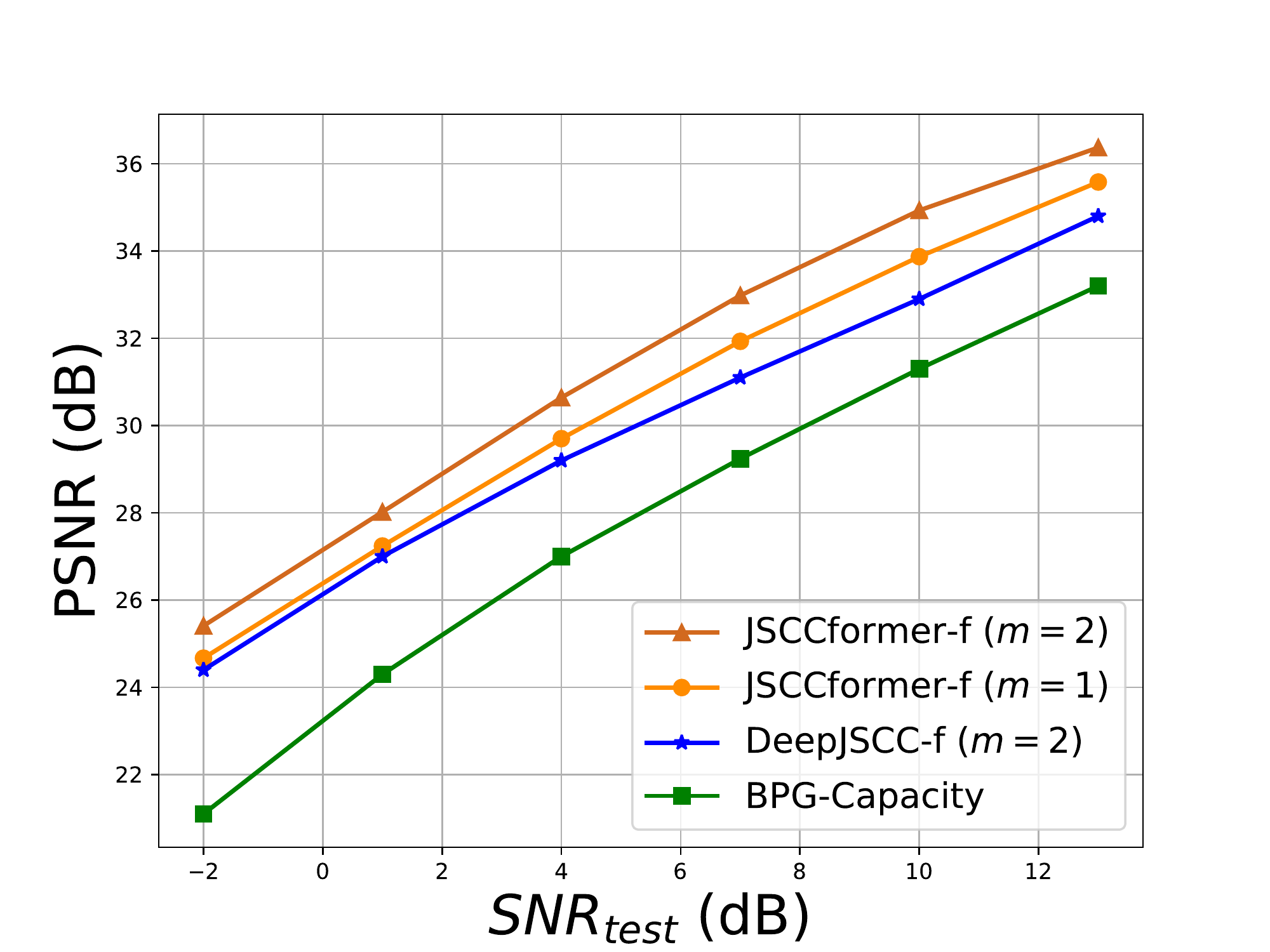}
    \end{minipage}%
    }%
    \subfloat[$R=1/3$]{
    \label{ViT_16} 
    \begin{minipage}[t]{0.32\linewidth}
    \centering
    \includegraphics[scale=0.3]{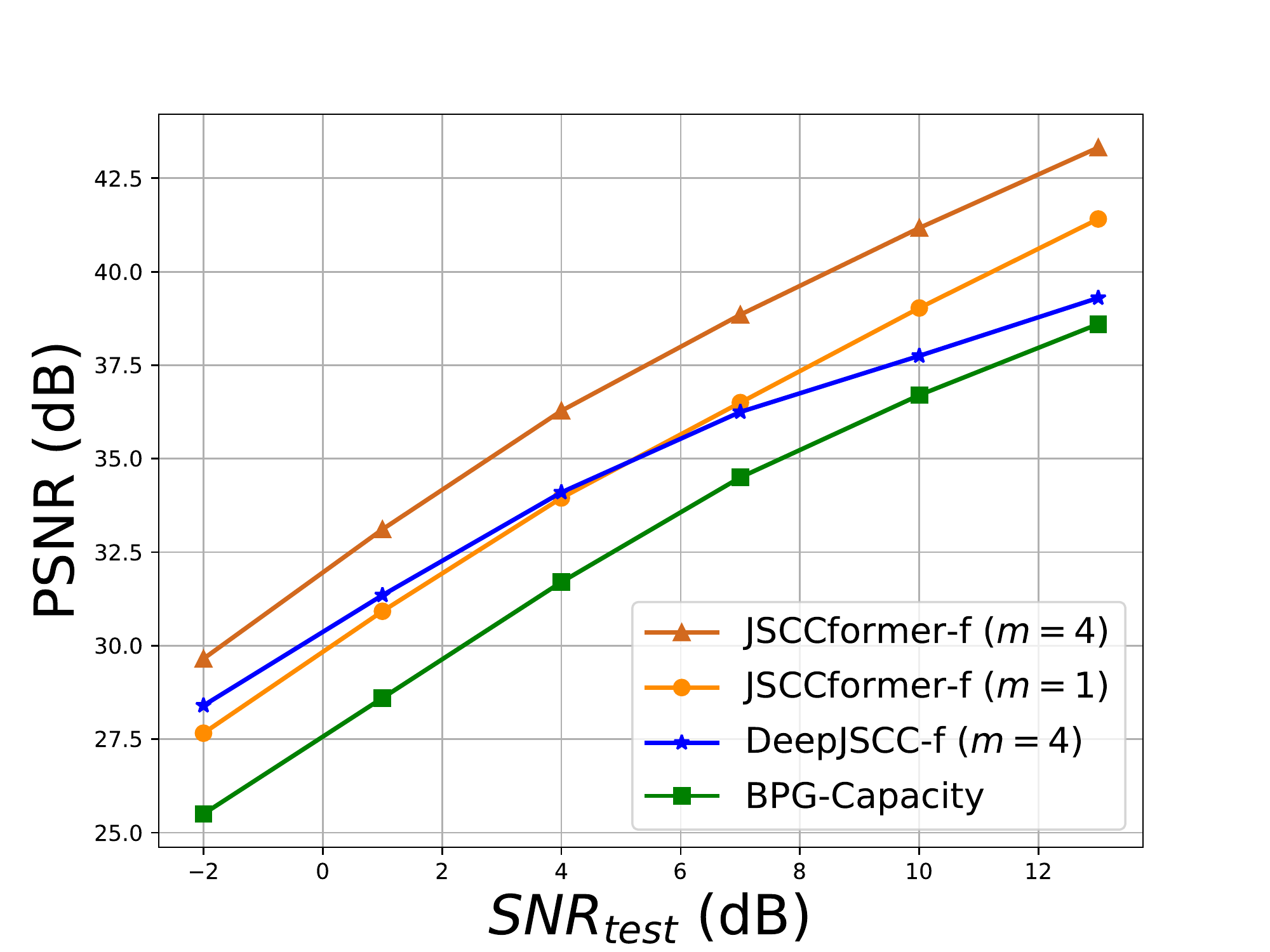}
    \end{minipage}%
    }%
    \subfloat[$R=1/6$]{
    \label{Vitf_8_snr_fad_performance} 
    \begin{minipage}[t]{0.32\linewidth}
    \centering
    \includegraphics[scale=0.3]{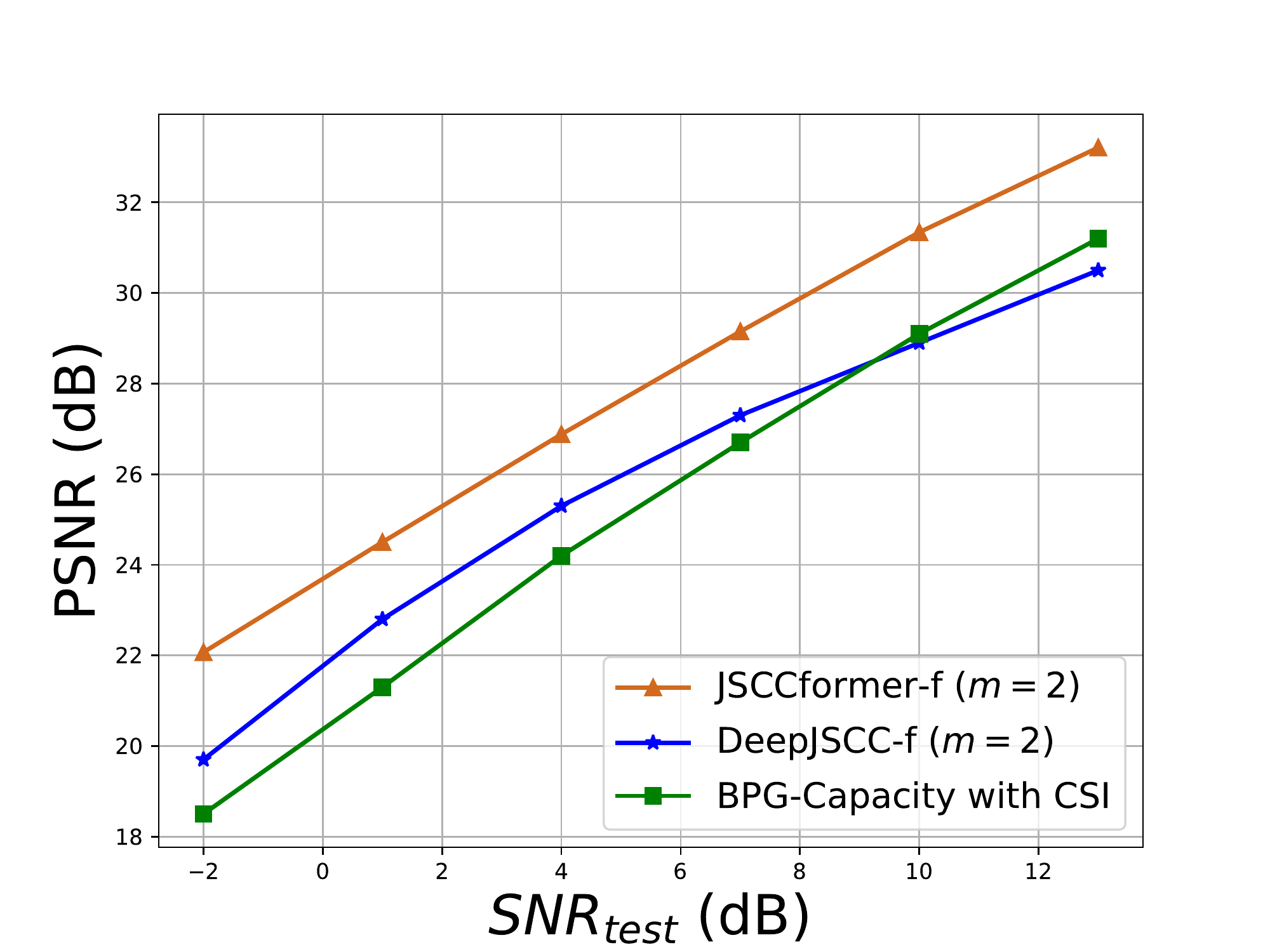}
    \end{minipage}%
    }%
    \centering
    \caption{ {Performance of different schemes at various SNR values and bandwidth ratios with noiseless feedback, where models in subfigures (a) and (b) undergo assessment within the AWGN channel, and models in subfigure (c) are  evaluated in a Rayleigh Fading channel.}}
    \label{ViT_JSCC} 
\end{figure*}

\section{Training and evaluation}
This section presents experimental results to evaluate the performance of our JSCCformer-f model across various scenarios. 
DeepJSCC-f \cite{kurka2020deepjscc} and the bounds on the conventional separation-based digital schemes, assuming capacity-achieving channel codes, are employed as benchmarks. In the conventional separation-based digital schemes, we use Better Portable Graphics (BPG) as the compression method and assume capacity-achieving channel codes, denoted as the BPG-Capacity scheme. It is noteworthy that the BPG-Capacity scheme is practically unattainable and requires perfect CSI at both the transmitter and receiver ends.

Unless stated otherwise, all experiments in this section were performed on the CIFAR10 dataset\cite{krizhevsky2009learning}, consisting of a training dataset of $50000$ images and a test dataset of $10000$ images, with the shape of $3\times 32\times 32$ (color, height, width). To achieve the optimal result, models of JSCCformer-f and DeepJSCC-f are all trained at a fixed average channel SNR value and tested at the same SNR value. All models were implemented in Pytorch with two GTX 3090Ti GPUs and optimized by an Adam optimizer. We use a learning rate of $0.00005$ and a batch size of $128$. Models were trained until the performance of a validation set stopped improving. Considering the model complexity and experiment performance, we set the $p=8$, $l=64$, and $c=48$ for image vectorization. We set $d=256$, $L_t=8$, and $N_s=8$ for each transformer layer of the ViT. 



\subsection{Transmission performance}
\subsubsection{General performance}
Considering AWGN channels, Figs. \ref{ViT_8} and \ref{ViT_16} present the PSNR performance versus different forward channel SNR values from $-2$dB to $13$dB, where the bandwidth ratio is set to $R=1/6$ and $R=1/3$, respectively. We assume the feedback link is noiseless for now, and the noisy feedback case will be considered later in Section \ref{noisy_section}.

We first evaluate JSCCformer-f without channel feedback, i.e., $m=1$. As shown in Figs. \ref{ViT_8} and \ref{ViT_16}, JSCCformer-f with $m=1$ significantly outperforms the BPG-Capacity scheme in all SNRs and bandwidth ratios. We note again that the BPG-Capacity performance is not achievable in the short block length regime. 
We next compare JSCCformer-f ($m=1$) with DeepJSCC-f ($m=2$).
As shown in Fig. \ref{ViT_8}, JSCCformer-f ($m=1$) can surprisingly outperform DeepJSCC-f ($m=2$) (up to $0.97$dB) at all SNR values, even though DeepJSCC-f utilizes additional feedback from the receiver. 

Then, we consider JSCCformer-f with $m>1$. For a fair comparison with the simulation results in \cite{kurka2020deepjscc}, we use the same number of block for JSCCformer-f. Specifically, we set $m=2$ for $R=1/6$ in Fig. \ref{ViT_8} and $m=4$ for $R=1/3$ in Fig. \ref{ViT_16}.
JSCCformer-f ($m>1$) improves the performance of JSCCformer-f ($m=1$) in all SNR values (up to $1.06$ dB for $m=2$ and $2.11$ dB for $m=4$). Further, it can be observed that a larger bandwidth ratio leads to a larger gain. We also observe that with the same number of block, JSCCformer-f significantly outperforms DeepJSCC-f at all test SNRs. The gains are up to $2.03$ dB and $3.88 dB$ for $R=1/6$ and $R=1/3$, respectively, demonstrating that JSCCformer-f can exploit feedback better than DeepJSCC-f. Moreover, with the increase in the bandwidth ratio, the improvement from JSCCformer-f becomes even more significant.

Fig. \ref{Vitf_8_snr_fad_performance} shows the performance of JSCCformer-f and DeepJSCC-f in fading channels, where the fading gain is sampled from $h\sim \mathcal{CN}(0,1)$. We observe that DeepJSCC-f outperforms the BPG-Capacity scheme only in the low SNR regime ($-2$dB to $10$dB), although we should remind that this is a highly favorable bound for the separation-based scheme. However, JSCCformer-f consistently outperforms both DeepJSCC-f and BPG-Capacity schemes in all SNR regimes with a significant gain of up to $2.71$dB and $3.57$dB, respectively. 

\begin{figure}[t]
\centering
    \includegraphics[scale=0.4]{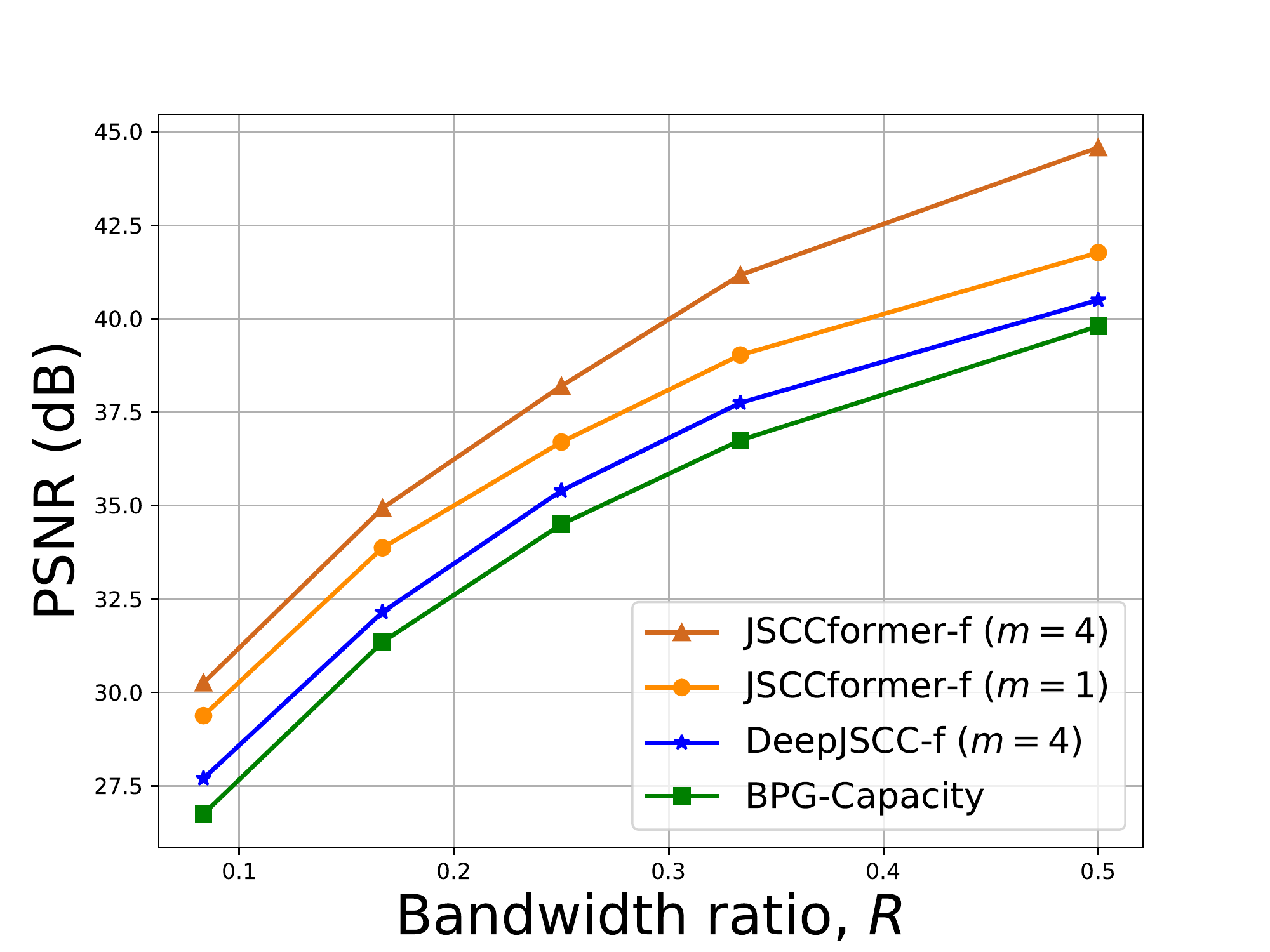}
\caption{Performance comparison of different models versus bandwidth ratio $R$ in  {AWGN channel} when SNR$=10$ dB with noiseless feedback.}
\label{Vitf_ratio}
\end{figure}

\subsubsection{Impacts of bandwidth ratio and block number}
We evaluate the performance of JSCCformer-f over a wide range of bandwidth ratios $R$ in Fig. \ref{Vitf_ratio}, where the forward SNR is fixed to $10$dB, the feedback is noiseless, and $m=4$. JSCCformer-f outperforms DeepJSCC-f in all bandwidth ratio settings with a gain of up to $3.42$dB. 
Interestingly, even with no feedback, JSCCformer-f ($m=1$) outperforms DeepJSCC-f $(m=4)$ in the low bandwidth ratio regime ($R\leq 0.45$).

\begin{figure}[t]
\centering
    \includegraphics[scale=0.4]{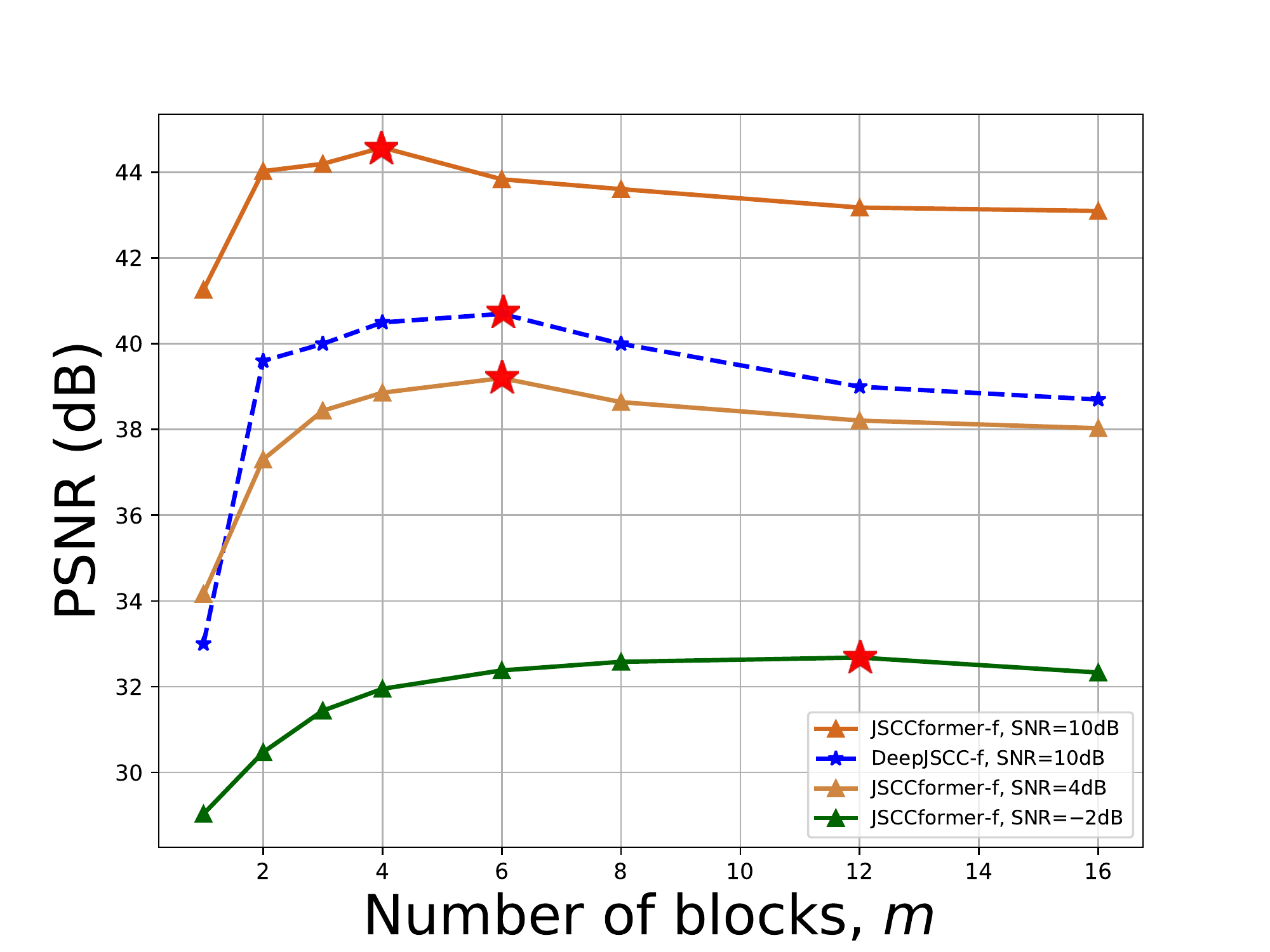}
    \caption{ {Performance of models over different block numbers and SNRs in AWGN channel with $R=1/2$, where the optimal block number is highlighted with a red star.}}
    \label{PSNR_over_iter_24}
\end{figure}

Next, we assess the impact of the block number on the PSNR of DeepJSCC-f and JSCCformer-f. To this end, we consider  {different} feedforward SNR values  {($-2$dB, $4$dB, and $10$dB)}, noiseless feedback, $R=1/2$, and vary the number of blocks $m$, from $1$ to  {$16$}. We note that, in principle, increasing $m$ means that we benefit from more feedback at the encoder. However, it can be observed in Fig. \ref{PSNR_over_iter_24} that the performances of JSCCformer-f and DeepJSCC-f do not always improve with $m$. Specifically, increasing $m$ brings some improvements at first for both DeepJSCC-f and JSCCformer-f; but the PSNR declines beyond a certain threshold. One possible explanation for this is the increasing learning difficulty brought about by increasing $m$.  {We can observe that the optimal $m$ values for JSCCformer-f are $m=4, 6, 12$ corresponding to  $SNR = 10$dB, $4$dB, $-2$dB, respectively. Notably, the optimal number of blocks tends to be larger under worse channel conditions. This can be explained by the fact that a greater number of feedback blocks can provide more information about the decoder's current beliefs and the channel conditions, which is more beneficial for the model in poor channel conditions.} 

 {We also remark that DeepJSCC-f adopts multiple encoders and decoders as different refinement layers; and hence unlike JSCCformer-f, DeepJSCC-f model complexity increases significantly with $m$.} In contrast, {JSCCformer-f} employs a unified pair of encoder and decoder, yielding less complexity while achieving better performance.  {Interestingly, both DeepJSCC-f and JSCCformer-f tend to converge towards a sub-optimal performance as $m$ becomes sufficiently large, thus revealing the existence of the trade-off between the training complexity and model performance.}  {Our experimental results suggest that $m=4$ provides a reasonable performance and complexity trade-off for JSCCformer-f in this channel condition, which can achieve competitive performance with an acceptable computation complexity.} 

\begin{figure}[t]
\centering
    \includegraphics[scale=0.4]{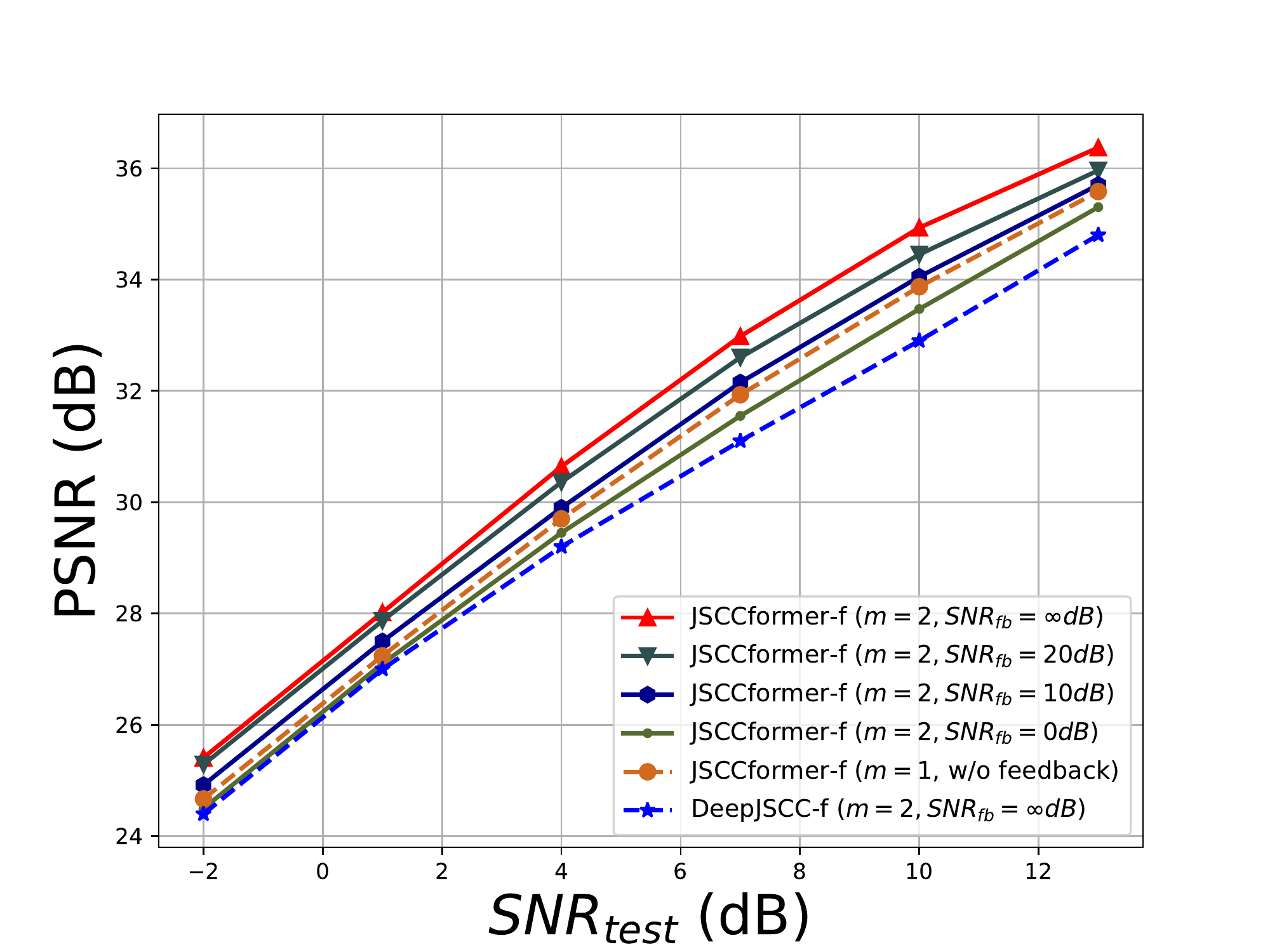}
    \caption{Performance of JSCCformer-f over noisy feedback link in AWGN channel, where $R=1/6$, $m=2$.}
    \label{ViTf_noisy_fb}
\end{figure}

\subsubsection{Noisy feedback channel}
\label{noisy_section}
In the above experiments, we have assumed a noiseless feedback channel. In this subsection, we investigate the impact of noisy feedback links for JSCCformer-f. Introducing noise in the feedback channel can lead to a drastic performance decrease\cite{kim2018deepcode} or even system breaking down\cite{kailath1967application}, as feedback can be dominated by noise. DeepJSCC-f is the first practical image transmission scheme that considers noisy feedback, which we shall use as a benchmark.

We repeat the experiments in Fig \ref{ViT_8} by replacing the noiseless feedback channel with an AWGN channel. The feedback channel SNR is set to $\text{SNR}_{\text{fb}}=0$dB, $10$dB, and $20$dB.
Fig. \ref{ViTf_noisy_fb} shows the performance of the JSCCformer-f model with different $\text{SNR}_{\text{fb}}$ values.
When $\text{SNR}_{\text{fb}}=20$dB (high-quality feedback), the performance only degrades slightly (for at most $0.48$dB), when compared with the perfect feedback case. When the feedback channel quality gets worse, e.g., $\text{SNR}_{\text{fb}}=10$dB, the PSNR continues to decrease, but the gap is less than $0.88$dB compared with the noiseless case).
This verifies the robustness of JSCCformer-f to feedback channel noise.
In particular, it can be observed that the performances of JSCCformer-f $(m=2)$ with $\text{SNR}_{\text{fb}}=20$dB and $\text{SNR}_{\text{fb}}=10$dB are still better than JSCCformer-f $(m=1)$ without feedback, indicating that JSCCformer-f can still make good use of the noisy feedback.
Even with larger feedback noise $\text{SNR}_{\text{fb}}=0$dB, our model can still outperform DeepJSCC-f with a perfect feedback link. It again demonstrates the superiority of our scheme in noisy feedback scenarios. 
Overall, we conclude that JSCCformer-f is robust to feedback noise.

\begin{figure}[t]
\centering
\centering
    \includegraphics[scale=0.4]{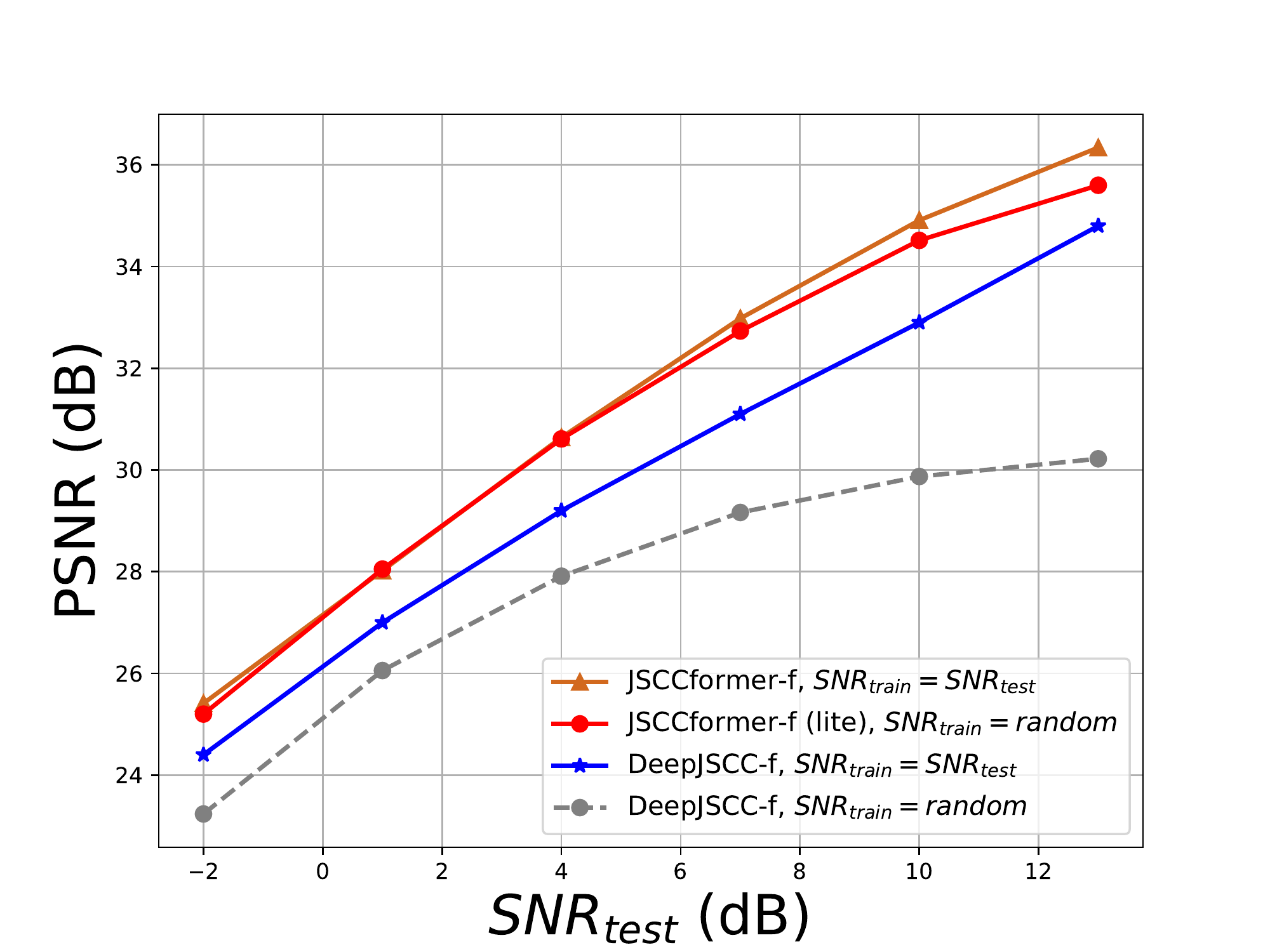}
    \caption{Performance of JSCCformer-f (lite) compared with other schemes in AWGN channel, where $R=1/6$, and $m=2$.}
    \label{adViT_snr_performance}
\end{figure}
\subsection{{Adaptability}}
This section validates the SNR-adaptability of JSCCformer-f. To verify the SNR adaptability, we adopt a random-SNR training method. That is, the SNRs in the training phase are uniformly sampled from the range $[-2,15]$ dB. After training, the well-trained model is then evaluated at different test SNRs. We expect the JSCCformer-f and JSCCformer-f (lite) encoder to learn the channel conditions from the feedback signal to help the model map the source signal into channel symbols via the SA mechanism. 

 {We first conduct an ablation study over different training strategies for JSCCformer-f and JSCCformer-f (lite) in Table \ref{ablation_over_training}, where $m=4$ and $R=1/3$. When trained for a specific SNR, JSCCformer-f achieves a superior performance, tailoring model parameters for this channel condition. On the other hand, when trained over random channel SNRs, both models experience a performance degradation. However, JSCCformer-f (lite) exhibits a reduced level of performance degradation, which is due to its simplified architecture, particularly the lack of a decoder module at the transmitter, which is difficult to adapt to varying channel conditions. These results highlight that JSCCformer-f (lite) is tailored for varying channel conditions, and we set the model trained for varying channel SNRs as the default JSCCformer-f (lite) model.}

 {To compare with the benchmark,} we repeat the experiments in Fig. \ref{ViT_JSCC} and plot the performance of JSCCformer-f, JSCCformer-f (lite) and DeepJSCC-f with different training strategies in Fig. \ref{adViT_snr_performance}. It can be observed that the DeepJSCC-f scheme exhibits a significant degradation in performance when trained with the random SNR strategy, indicating its inability to adapt to diverse channel conditions. Then, we compare the performance of the JSCCformer-f (lite) model trained using random SNRs and the optimal DeepJSCC-f models trained at each individual SNR. As shown, JSCCformer-f (lite) outperforms DeepJSCC-f across all SNR scenarios, with a maximum gain of $1.64$ dB. {JSCCformer-f} (lite) can adapt to different test SNRs.  {Importantly, this channel adaptability is attributed to the proposed method's inherent attention mechanism rather than being solely an outcome of the random training strategy.} 

\begin{figure}[t]
    \centering
    \centering
    \includegraphics[scale=0.4]{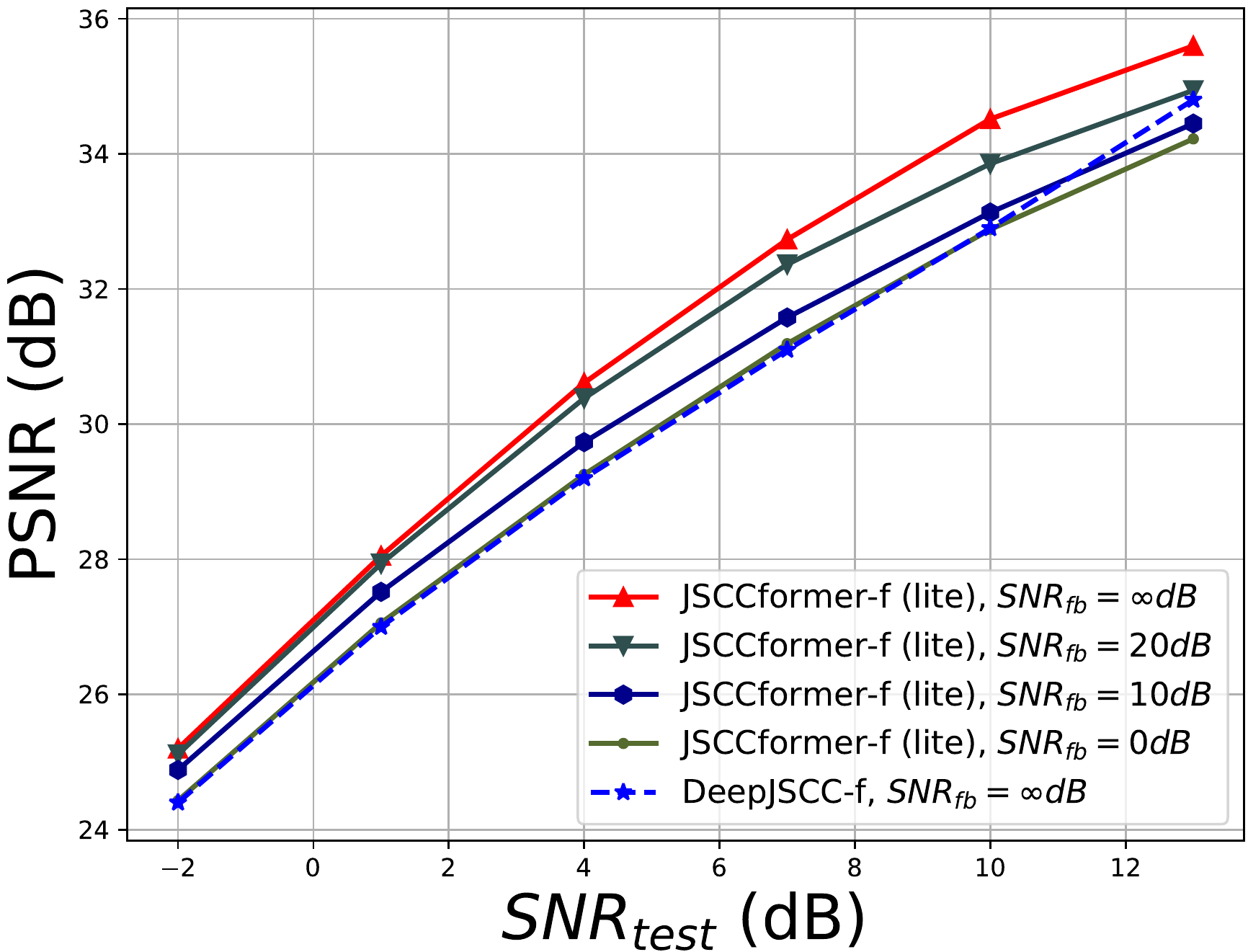}
    \caption{\textcolor{blue}{Performance of JSCCformer-f (lite) trained with random SNRs over noisy feedback link in the AWGN channel, where $R=1/6$ and $m=2$.}}
    \label{Ad_ViT_fb_noise}
\end{figure}

\begin{table}[t]
    \caption{ {Ablation study over different training strategies (fixed SNR or random SNR), where $m=4$ and $R=1/3$.}}
    \centering
    \begin{tabular}{c|c|c|c}
    \textbf{Different channel SNR } & $1$dB & $7$dB&$13$dB\\
    \hline
      JSCCformer-f (fixed)& \textbf{$\bm{33.11}$dB}& \textbf{$\bm{38.85}$dB} & \textbf{$\bm{43.32}$dB}\\
      JSCCformer-f (random) & ${32.69}$dB & ${38.31}$dB&  ${41.47}$dB\\
      \hline
     JSCCformer-f (lite) (fixed)& \textbf{$\bm{32.91}$dB}& \textbf{$\bm{38.64}$dB}&  \textbf{$\bm{42.77}$dB} \\
    JSCCformer-f (lite) (random)&  ${32.89}$dB & $38.57$dB& ${41.87}$dB \\
    \end{tabular}
    \label{ablation_over_training}
\end{table}

\begin{table*}[t]
\caption{ {The number of parameters, FLOPs, and coding time for different models; bold figures correspond to the minimal values among different models for each $m$ value.}}
\begin{center}
\begin{tabular*}{1\textwidth}{@{\extracolsep{\fill}}|c|c|ccccccc|}
\hline 
\multicolumn{2}{|c|}{\textbf{Block number $m$}}&1&2&3&4&6&8&{12} \\
\hline 
\multirow{3}{*}{\textbf{Parameters} (millions)}&\text{DeepJSCC-f}& \textbf{10.58}& 25.4&40.25&55.21&85.10&114.99&{174.76}\\
&\text{JSCCformer-f}& 12.93&12.95& 12.96&12.97&12.99&13.02&{13.07}\\
		&\text{JSCCformer-f ({lite})}& 12.93& \textbf{12.93}&\textbf{12.93}&\textbf{12.93}&\textbf{12.93}&\textbf{12.93}&{\textbf{12.93}}\\
\hline 
\multirow{3}{*}{\textbf{FLOPs} (G)}
&\text{DeepJSCC-f}&1.06& 2.82&4.58&6.34&9.86&13.38&{20.42}\\
&\text{JSCCformer-f}& 0.83& 1.67&2.51&3.35&5.03&6.72&{10.12}\\
&\text{JSCCformer-f (lite)}&\textbf{0.83}&\textbf{1.24}&\textbf{1.64}&\textbf{2.05}&\textbf{2.86}&\textbf{3.66}&{\textbf{5.28}}\\
\hline
\hline

\multirow{2}{*}{\textbf{{Encoding time}} (ms)}
		&\text{{DeepJSCC-f}}& \textbf{1.82}& \textbf{6.12}&\textbf{10.03}&\textbf{14.10}&\textbf{21.99}&\textbf{32.07}&\textbf{45.94}\\

		&\text{{JSCCformer-f}}& 4.75& 13.76&23.20&32.64&50.22&71.54&111.02\\
		&\text{{JSCCformer-f} (lite)}& 4.54&9.04&13.72&18.20&27.64&36.83&58.82\\
    \hline 
\multirow{2}{*}{\textbf{{Decoding time}} (ms)}
		&\text{{DeepJSCC-f}}& \textbf{1.84}& 4.32&6.13&8.21&12.15&17.40&24.12\\

		&\text{{JSCCformer-f}}& 4.68& 5.44&\textbf{4.89}&5.36&4.96&\textbf{3.78}&4.46\\
		&\text{{JSCCformer-f} (lite)}& 4.97&\textbf{4.12}&5.00&\textbf{4.73}&\textbf{4.79}&3.98&\textbf{4.15}\\
  \hline
\multirow{2}{*}{\textbf{{Total transmission time}} (ms)}
		&\text{{DeepJSCC-f}}& \textbf{3.66}& \textbf{10.44}&\textbf{16.16}&\textbf{22.31}&34.14&49.47&70.06\\

		&\text{{JSCCformer-f}}& 9.43& 19.02&{28.09}&38.00&55.81&{75.32}&115.48\\
		&\text{{JSCCformer-f} (lite)}& 8.51&{13.16}&18.72&{22.93}&\textbf{32.43}&\textbf{40.81}&\textbf{62.97}\\
  
    \hline 
\end{tabular*}
\vspace{-5pt}
\label{table_para}
\end{center}
\end{table*}


{In a comparative analysis between JSCCformer-f (lite) and JSCCformer-f from Fig. \ref{adViT_snr_performance}}, the results indicate that the performance of JSCCformer-f (lite) is only slightly worse than the optimal performance of JSCCformer-f at all SNRs. The performance gap is smaller than $0.74$dB. In summary, our JSCCformer-f (lite) is able to achieve close performance to JSCCformer-f in a wide range of SNRs and is significantly better than DeepJSCC-f. It is worth noting that we do not feed the channel SNRs explicitly to the encoder and decoder. Instead, they are implicitly learned by JSCCformer-f from the observed channel outputs. 
 
We also evaluate the robustness of JSCCformer-f (lite) to the noisy feedback case while maintaining channel adaptability.
We consider JSCCformer-f (lite) with noisy AWGN feedback links with feedback SNRs of $0$dB, $10$dB, and $20$dB, whose performance is illustrated in Fig. \ref{Ad_ViT_fb_noise}. When $\text{SNR}_{\text{fb}}=20$dB, the performance is only slightly worse (the gap is smaller than $0.6$dB) compared to that with a noiseless feedback channel. When $\text{SNR}_{\text{fb}}=10$dB and $\text{SNR}_{\text{fb}}=0$dB, the performance continues to degrade; the gap from the noiseless feedback case is up to $1.1$dB and $1.3$dB, respectively. 
Compared to DeepJSCC-f with ideal feedback, JSCCformer-f (lite)  exhibits superior performance even when the feedback channel SNR is $0$dB. This observation confirms the superiority of our JSCCformer-f (lite) model and its robustness to noisy feedback.
 
\subsection{Model efficiency}
This section evaluates the model efficiency of the different schemes. Table \ref{table_para} presents the model size, floating point of operations (FLOPs), and the coding time for DeepJSCC-f, JSCCformer-f, and JSCCformer-f (lite) with different number of blocks on the CIFAR10 dataset (with a batch size of $1$ and $R=1/2$).

As shown in the table, the parameter count of DeepJSCC-f rapidly increases as the block number increases, leading to a more complex model. Conversely, JSCCformer-f and JSCCformer-f (lite) maintain a lower parameter count, even as $m$ increases ($m>6$). Notably, JSCCformer-f (lite) exhibits the lowest parameter count when $m>1$. Interestingly, when $m=12$, JSCCformer-f (lite) achieves improved performance while reducing memory consumption by up to $92.6\%$ compared to DeepJSCC-f.  {In terms of computational complexity, it becomes evident that DeepJSCC-f incurs greater computational costs (FLOPs), particularly as the number of blocks increases.} JSCCformer-f (lite) requires up to $3.8$x times fewer FLOPs while maintaining competitive performance, indicating superior efficiency. Our JSCCformer-f-based methods demand fewer computations to achieve optimal performance, implying significantly lower latency in practical use.

 {To bring a more comprehensive understanding of the model's computation cost, we presented the encoding and decoding time of different methods with a GPU of RTX A6000 and a CPU of Intel Xeon Gold 5220R in Table \ref{table_para}. The coding speed of the DeepJSCC-f is the fastest, coming from the well-optimized CNNs inference library. For the JSCCformer-f, it is feasible to maintain the average coding time at an acceptable range within $51$ms when $m<6$. Compared with JSCCformer-f, JSCCformer-f (lite) significantly reduces the coding time, with potential savings of up to $47 \%$. Regarding the decoding time, it is noteworthy that the decoding operation for JSCCformer-based models is executed just once, with an average decoding time within $5$ms, indicating a rapid decoding speed at the receiver. Conversely, the DeepJSCC-f model must conduct multiple decoding operations at each block, resulting in high receiver-side latency, especially when $m$ increases.}

 {In summary, our proposed method demonstrates both practicality and efficiency in terms of the model size and computation complexity. It is worth highlighting that further optimization exists in our models' inference speed, encompassing alternative hardware options, different implementations, and acceleration techniques for ViT.} 

\subsection{{High resolution dataset and visualization}}
In order to evaluate the model generalizability, we validate the JSCCformer-f and JSCCformer-f (lite) with $m=2$ on the Kodak dataset using $SNR_{test}\in [1,10]$dB and $R=1/12$, utilizing CPE method. Specifically, models are trained with randomly cropped $256 \times 256$ patches from the ImageNet dataset and subsequently evaluated on the Kodak dataset. 

\begin{table}[t]
    \caption{ {Performance of models over the Kodak dataset.}}
    \centering
    \begin{tabular}{c|c|c|c|c}
    \textbf{Different models} & $-2$dB &$1$dB & $4$dB& $7$dB\\
    \hline
     JSCCformer-f & $\bm{29.2}$dB & $\bm{30.84}$dB & $\bm{32.05}$dB& $\bm{33.16}$dB\\
     JSCCformer-f (lite)& ${29.16}$dB & $30.76$dB & ${31.90}$dB & $32.91$dB \\
     DeepJSCC-f & ${28.02}$dB& $29.57$dB & ${31.04}$dB & $32.20$dB \\
    \end{tabular}
    \label{Kodak_vit_f}
\end{table}

\begin{figure*}[h]
\centering
    \includegraphics[scale=0.58]{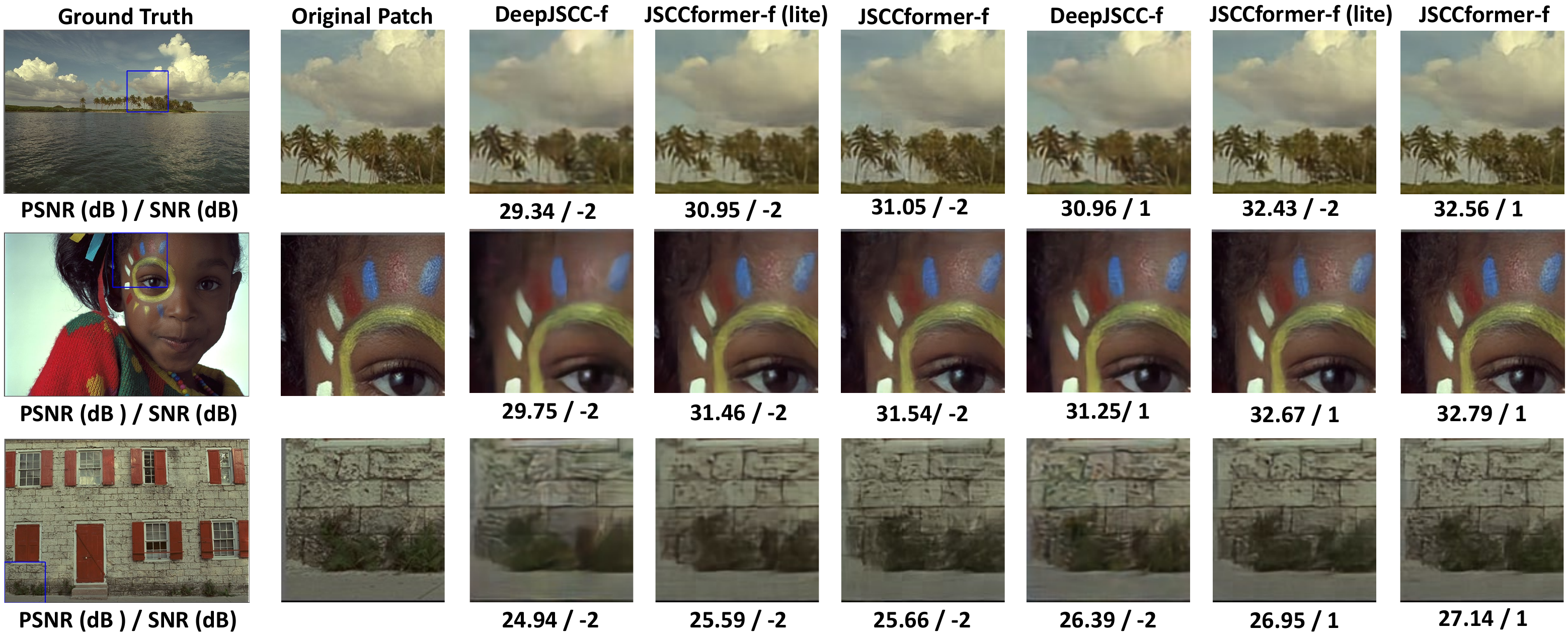}
    \caption{\textcolor{blue}{Visual comparisons of images transmitted by DeepJSCC-f, JSCCformer-f (lite), and JSCCformer-f over the AWGN channel, where the model is trained on the ImageNet dataset and validated on the Kodak dataset at various SNR values and $R=1/12$.}}
    \label{vis_vitf_patch}
\end{figure*}

\begin{figure*}[!h]
  \centering
  \begin{tabular}{ccc}
  \begin{minipage}[b]{0.6\columnwidth}
		\centering
		\raisebox{-.45\height}{\includegraphics[width=\linewidth]{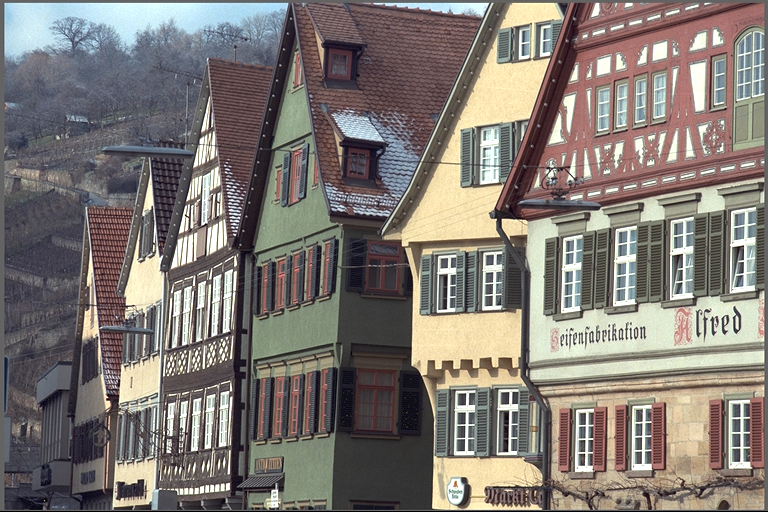}}
	\end{minipage}
    &      
    \begin{minipage}[b]{0.6\columnwidth}
		\centering
		\raisebox{-.45\height}{\includegraphics[width=\linewidth]{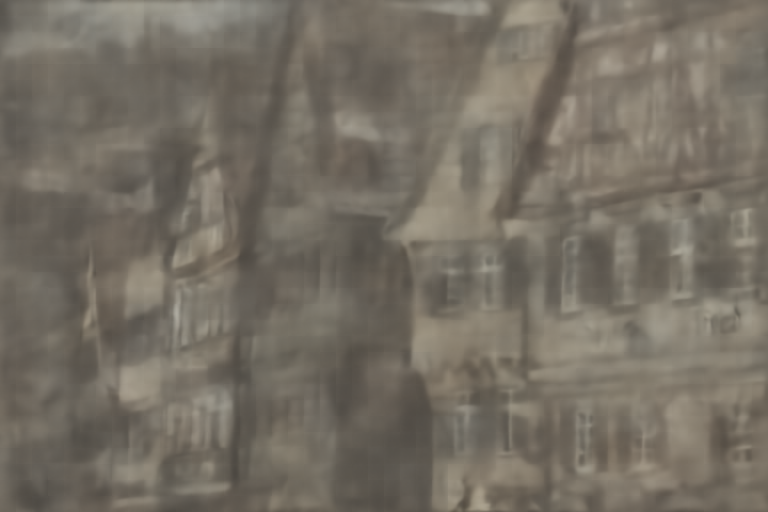}}
	\end{minipage}
     &
     \begin{minipage}[b]{0.6\columnwidth}
		\centering
		\raisebox{-.45\height}{\includegraphics[width=\linewidth]{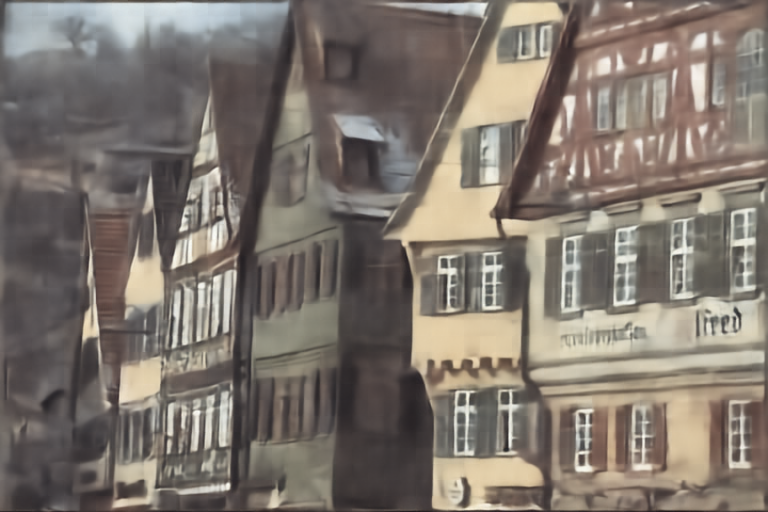}}
	\end{minipage}
	\\
     \small{Ground Truth} &
     \small{Reconstruction after the $1$-st interaction,}&
     \small{Reconstruction after the $2$-nd interaction,}\\
      &
     \small{PSNR=$13.98$dB}&
     \small{PSNR=$18.84$dB}\\
     \begin{minipage}[b]{0.6\columnwidth}
		\centering
		\raisebox{-.45\height}{\includegraphics[width=\linewidth]{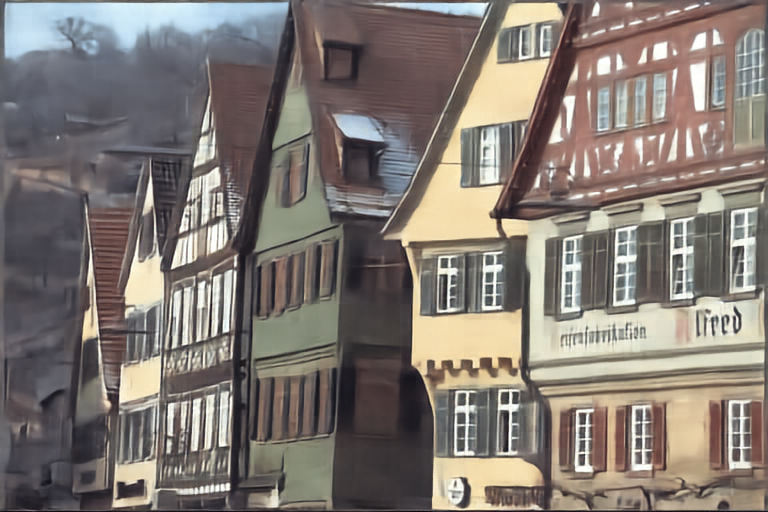}}
	\end{minipage} 
    &
               \begin{minipage}[b]{0.6\columnwidth}
		\centering
		\raisebox{-.45\height}{\includegraphics[width=\linewidth]{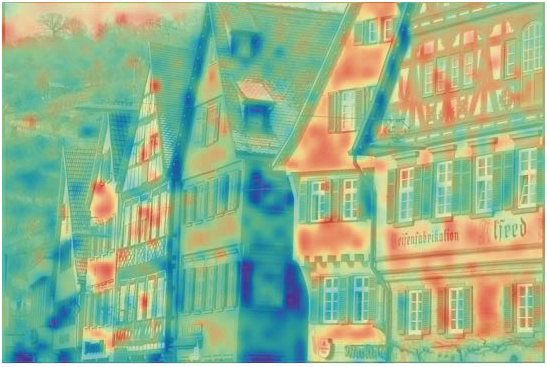}}
	\end{minipage}   
    &
       \begin{minipage}[b]{0.6\columnwidth}
		\centering
		\raisebox{-.45\height}{\includegraphics[width=\linewidth]{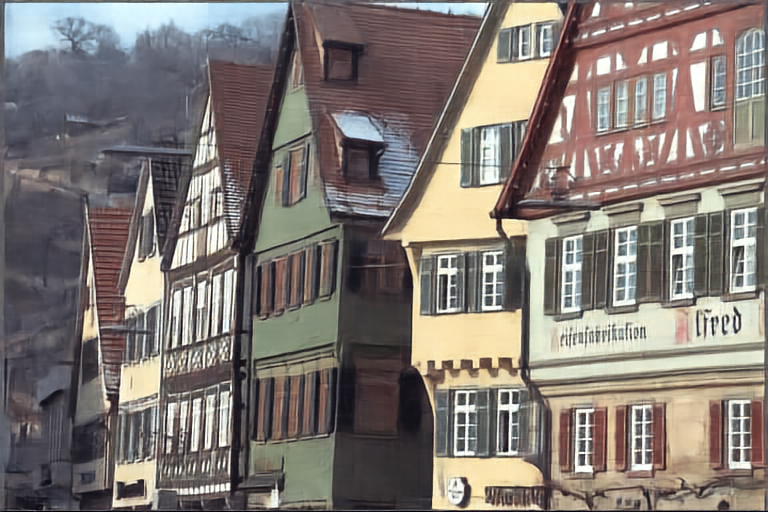}}
	\end{minipage}\\
\small{Reconstruction after the $3$-rd interaction,}&  
    \small{Attention map after the $3$-rd interaction} & 
    \small{Reconstruction after the $4$-th interaction,}\\
    \small{ PSNR=$22.24$dB}&  
    & 
    \small{PSNR=$24.15$dB}\\
  \end{tabular}
  \caption{ {Visualization of the reconstructed image with JSCCformer-f after each interaction for an input image from the Kodak dataset, where the model is trained on the ImageNet dataset with $R=1/12$, and $SNR=-2$dB.}}
    \label{vis_kodak}

\end{figure*}
The performances of JSCCformer-f, trained at specific SNRs, and JSCCformer-f (lite), trained with random SNRs, are compared against DeepJSCC-f over the AWGN channel, and the results are presented in Table. \ref{Kodak_vit_f}, where  $R=1/12$, $m=2$. In high-resolution datasets, JSCCformer-f based schemes maintain state-of-the-art performance and channel adaptability across various SNRs. Moreover, a sufficiently large dataset (such as ImageNet) allows our model to perform well on previously unseen images across a wide range of channel conditions (SNRs).

We visualize the comparisons of the recovered image patches by \textcolor{blue}{DeepJSCC-f, JSCCformer-f (lite) and JSCCformer-f} in Fig. \ref{vis_vitf_patch}. We can observe that JSCCformer-f performs the best with more detailed high-frequency features (e.g., hair and trees), particularly in the low SNR regime, e.g., at $-2$ and $1$ dB. \textcolor{blue}{JSCCformer-f (lite) demonstrates comparable visual performance to JSCCformer-f and exhibits a noteworthy improvement over DeepJSCC-f.}

\begin{figure*}[h]
\centering
    \includegraphics[scale=0.4]{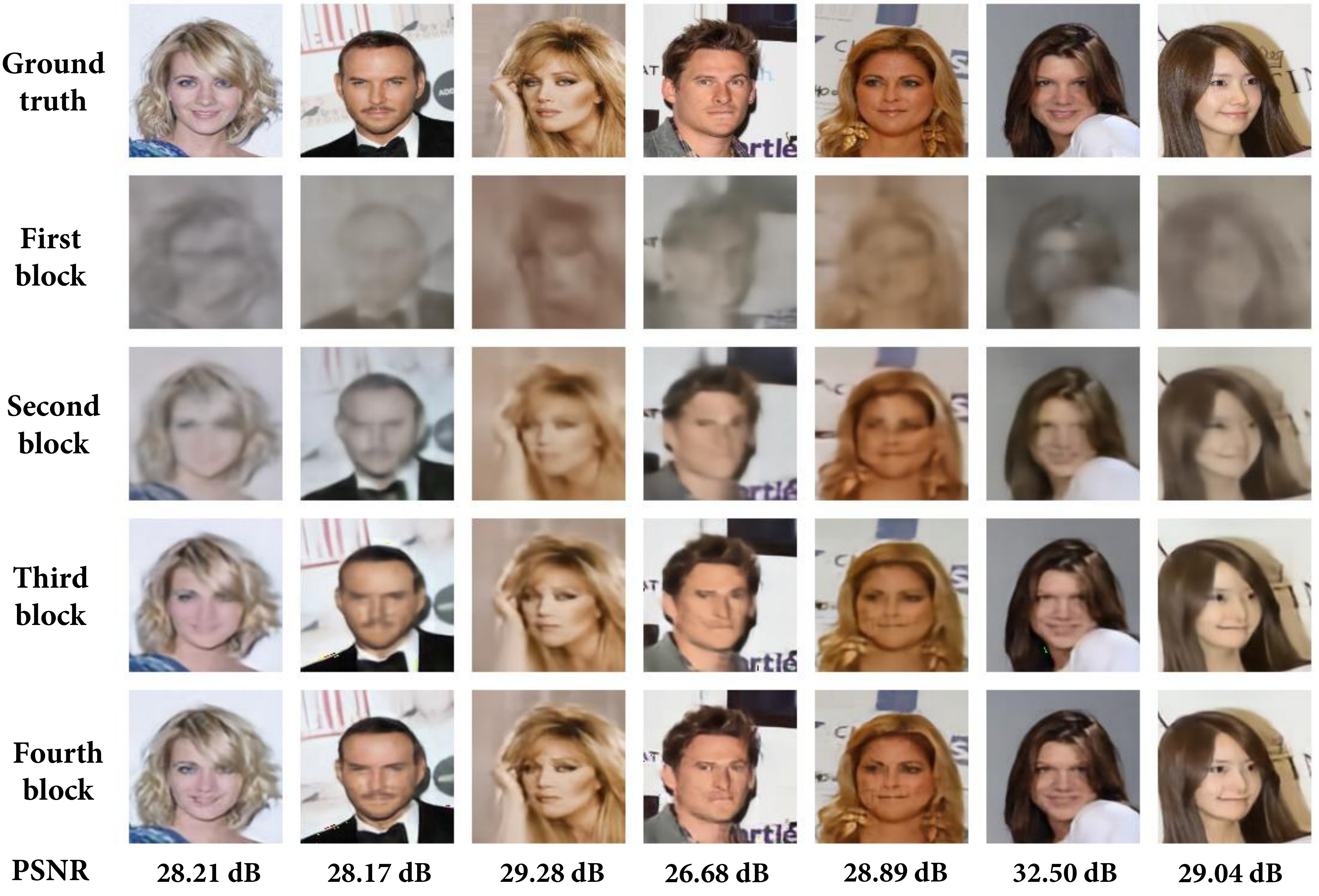}
  \caption{ {Visualization of the reconstructed image with JSCCformer-f after each interaction for images from the CelebA dataset, where the model is trained on the ImageNet dataset with $R=1/12$, and $SNR=-2$dB.}}
    \label{vis_celeba}
\end{figure*}

\begin{figure*}[h]
    \centering
    \subfloat[ {$R=1/6$ for CIFAR10 dataset.}]{
    \label{ViT_LPIPS_cifar} 
    \begin{minipage}[t]{0.48\linewidth}
    \centering
    \includegraphics[scale=0.4]{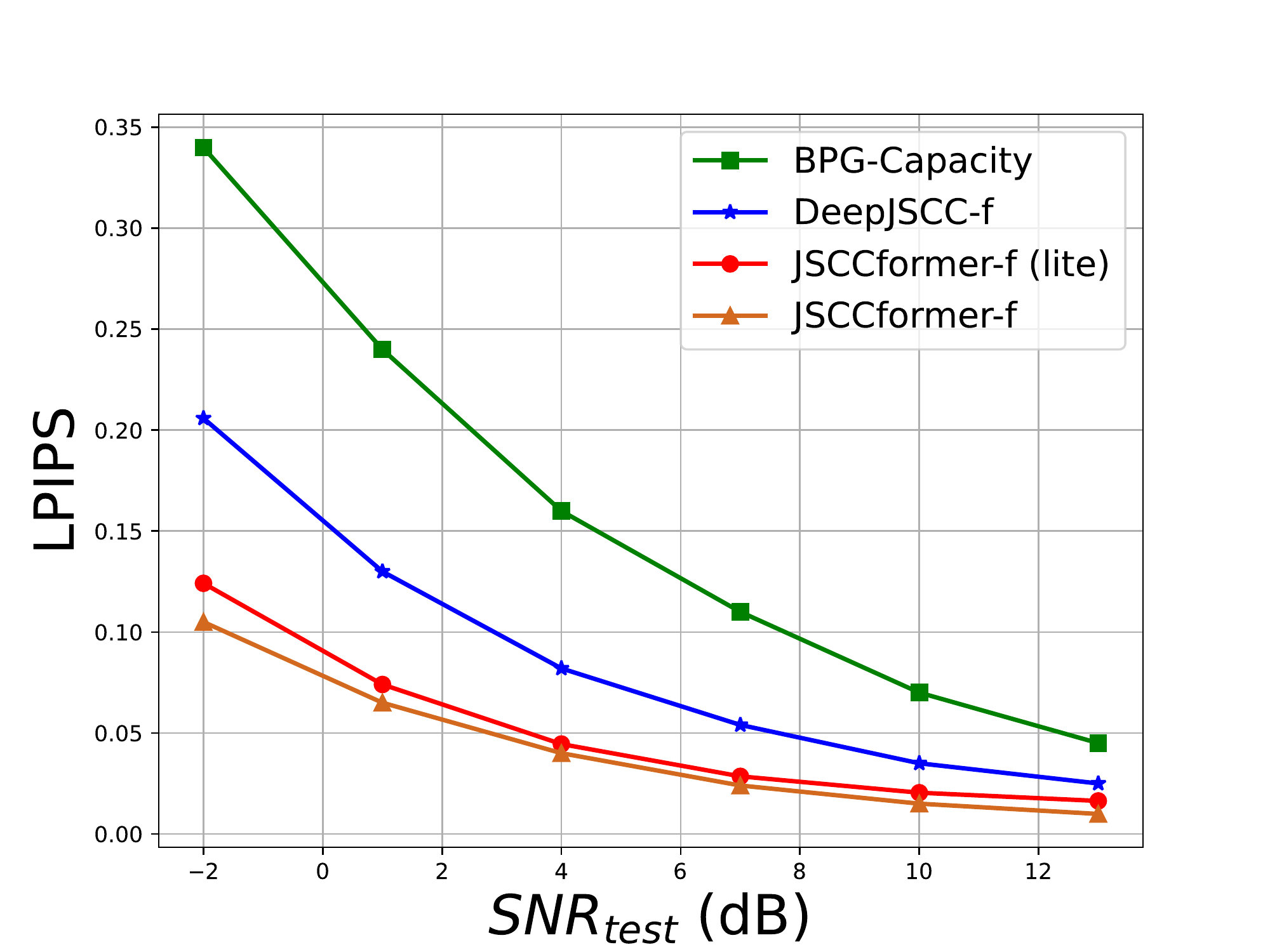}
    \end{minipage}%
    }%
    \hspace{-0.5mm}
    \subfloat[ {$R=1/12$ for CelebA dataset.}]{
    \label{ViT_LPIPS_celeba} 
    \begin{minipage}[t]{0.48\linewidth}
    \centering
    \includegraphics[scale=0.4]{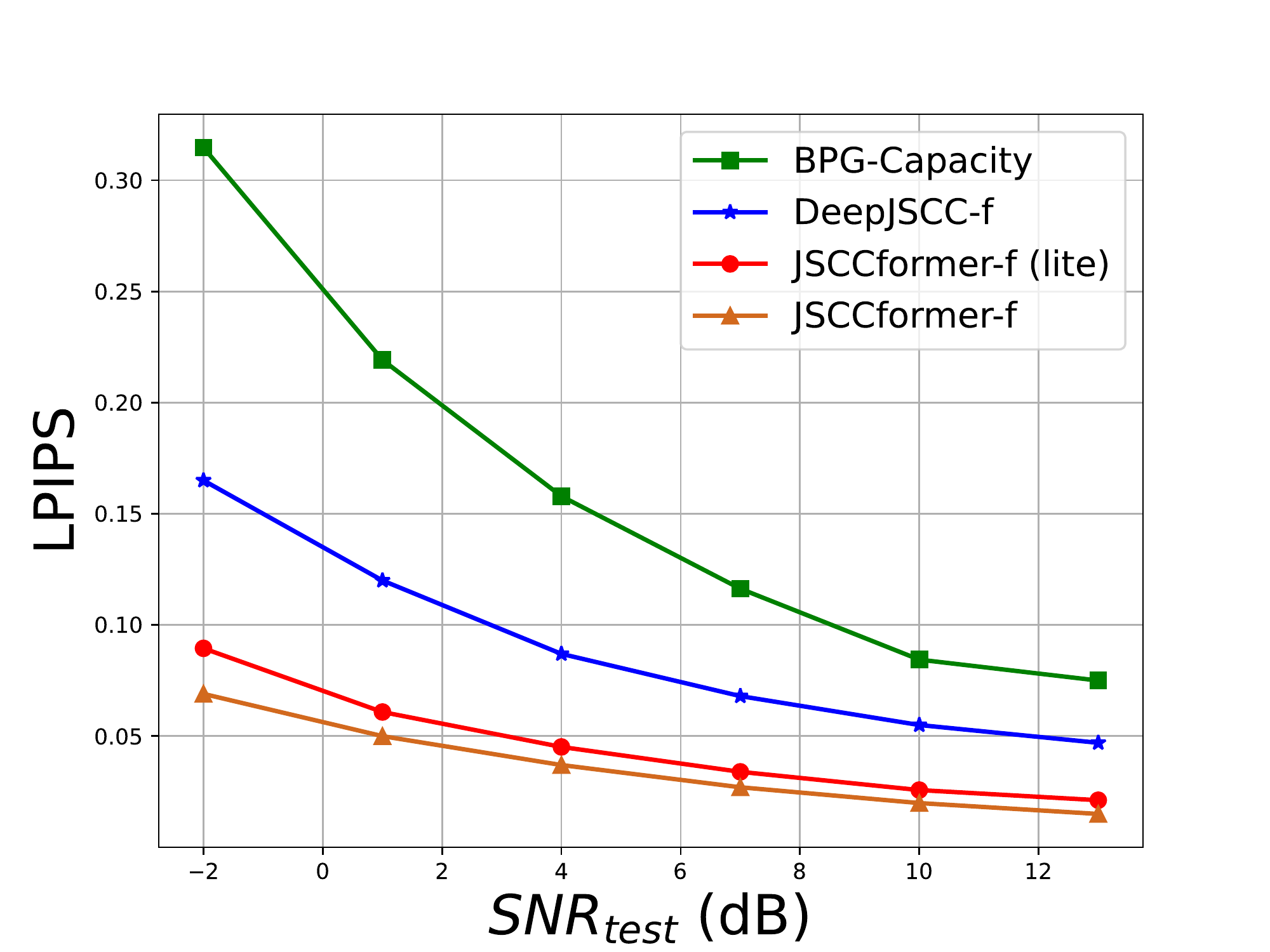}
    \end{minipage}%
    }%
    \centering
    \caption{ {LPIPS performance of different schemes, where models are trained with $m=2$.}}
    \label{ViT_JSCC_lpips} 
\end{figure*}

\begin{figure*}[tb]
    \centering
    \subfloat[]{
    \label{P_ratio} 
    \begin{minipage}[t]{0.48\linewidth}
    \centering
    \includegraphics[scale=0.4]{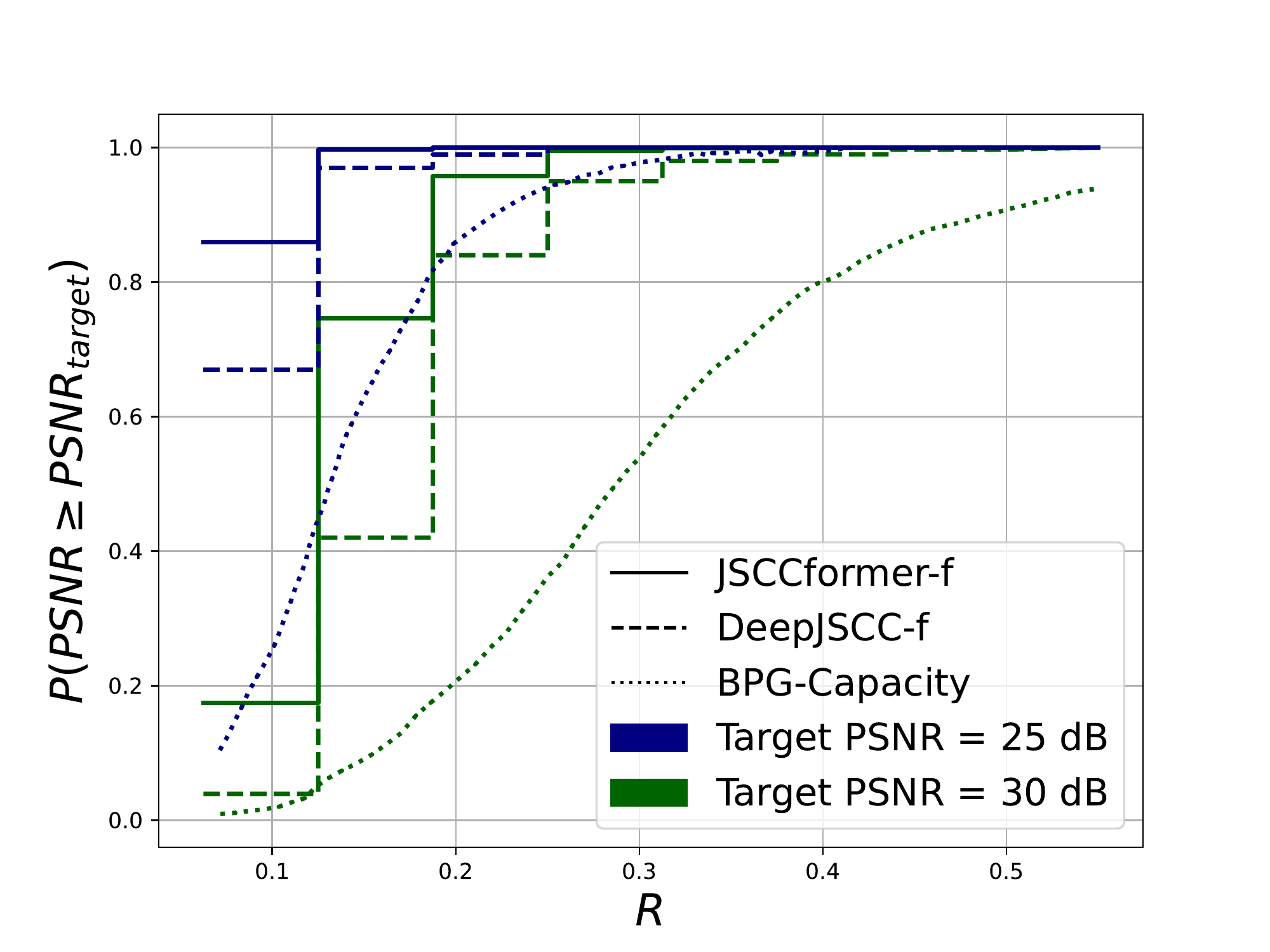}
    \end{minipage}%
    }%
    \hspace{-0.5mm}
    \subfloat[]{
    \label{average_ratio_PSNR} 
    \begin{minipage}[t]{0.48\linewidth}
    \centering
    \includegraphics[scale=0.4]{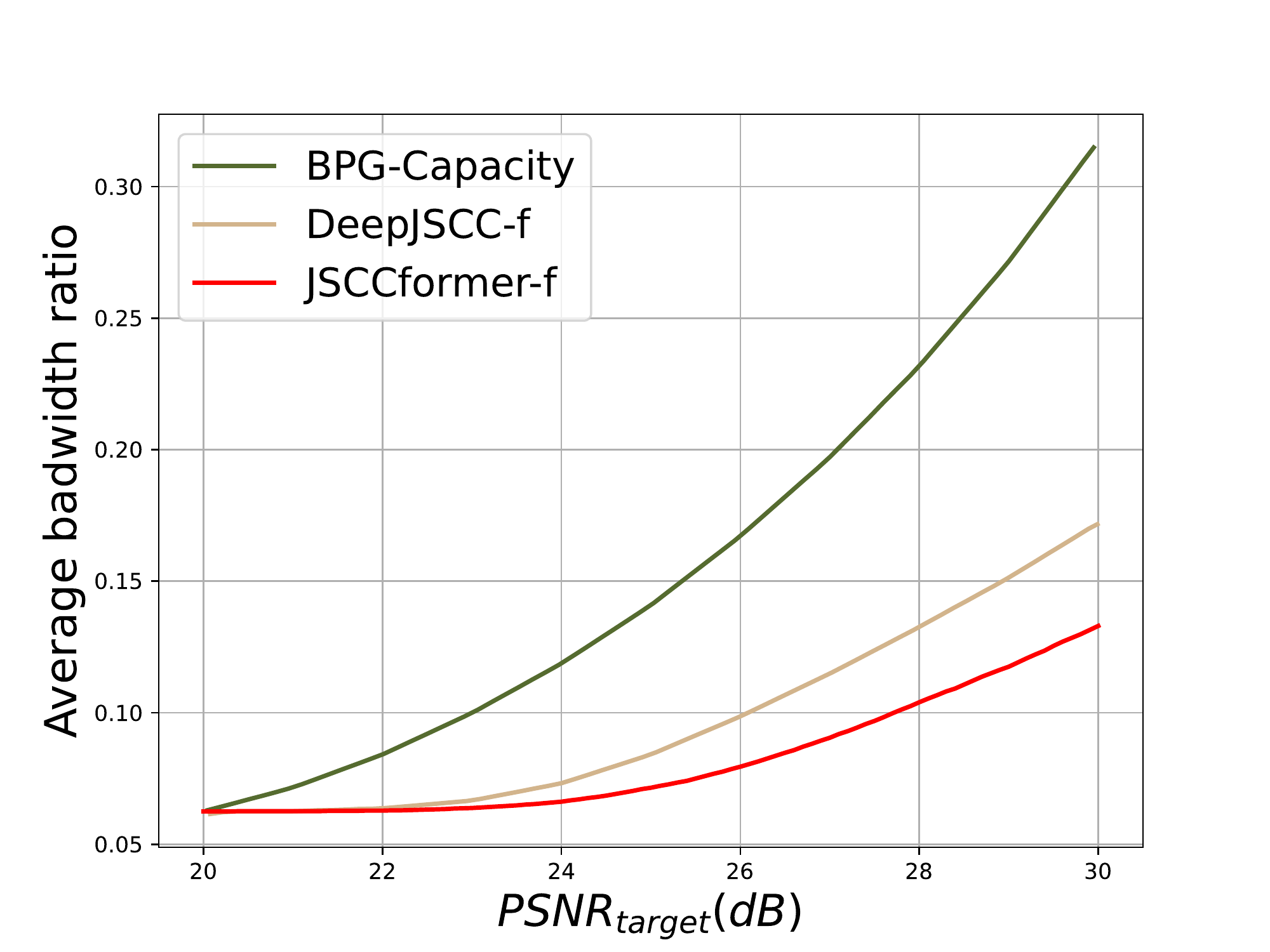}
    \end{minipage}%
    }%
    \centering
    \caption{Target PSNR analysis of different schemes for variable rate transmission. (a) The cumulative distribution function over various bandwidths given a specific target PSNR. (b) The average bandwidth ratio to achieve a specific target PSNR.}
    \label{Target_PSNR} 
\end{figure*}
Additionally, the effect of each interaction and the attention of the JSCCformer-f model in each transmission block are presented in Fig. \ref{vis_kodak} and \ref{vis_celeba},  {where we trained the models from the ImageNet dataset and tested the models over the Kodak and the CelebA datasets.}  {We can observe that the reconstruction performance is significantly improved after each interaction, particularly in some high-frequency details. Interestingly, the model tends to acquire an initial understanding of the overall semantic structure after the initial interaction, subsequently recovering a rough sketch. It then proceeds to enhance and refine its prior reconstructions through successive interactions. Furthermore, the attention map of the model in the final interaction revealed that our model could concentrate on some blurry areas and refine them in the next block, demonstrating how our method refines results in a coarse-to-fine manner based on the attention mechanism.}

\subsection{{Perceptual quality evaluation}}
 {To conduct a more comprehensive assessment of the model's perceptual qualities, we extend our evaluation with LPIPS measurements. To enhance the optimization of the LPIPS metric, akin to prior works\cite{erdemir2022generative}, we introduce an additional LPIPS term into the loss function, formulated as follows: $\mathcal{L}(\bm{\theta},\bm{\phi} )=\mathbb{E}\big[\|\bm{S}-\bm{\hat{S}}\|^2_2\big]+\lambda\cdot \text{LPIPS}(\bm{S},\bm{\hat{S}})$, where $\lambda$ is set as $0.1$ in our simulations.}

 {As shown in Fig. \ref{ViT_JSCC_lpips}, JSCCformer-f achieves the best perceptual performance across a spectrum of channel conditions, bandwidth ratios, and datasets. In particular, in poor SNR conditions, JSCCformer-f exhibits a substantial performance advantage over DeepJSCC-f and BPG-Capacity scheme, with a maximum LPIPS discrepancy of $0.235$ and $0.245$ at SNR $-2$dB for CIFAR10 and CelebA, separately. A similar trend is observed with the JSCCformer-f (lite), which preserves adaptability to channel variations but suffers a slight performance degradation relative to the original JSCCformer-f model.}

\subsection{Variable Rate Transmission}
The JSCC coding problem with channel feedback can be reformulated by imposing a particular transmission quality objective and minimizing the corresponding channel bandwidth, where more significant gains can be observed when considering variable-length coding\cite{kurka2020deepjscc,kostina2017joint}. As previously emphasized, the design of JSCCformer-f, which maps the channel symbols considering the decoder's present belief of the transmission, is inherently suitable for variable rate transmission. Specifically, the transmitter can determine the stopping time by analyzing the decoder's current knowledge of the transmission with the help of perfect channel output feedback. The transmitter stops the transmission of the remaining blocks when the current performance meets the desired target.

We evaluate the performance of JSCCformer-f with variable transmission rate, where we compute the average bandwidth necessary to achieve a predefined PSNR target. For a fair comparison, we set $m=8$, same with DeepJSCC-f. In the BPG-Capacity scheme, we compute the number of channel uses required to transmit the minimum number of compressed bits while satisfying the target PSNR value using a capacity-achieving channel code. In order to improve the efficacy of variable rate transmission, we fine-tune the well-trained JSCCformer-f while preserving the end-to-end transmission performance, which is achieved through the utilization of the subsequent loss function:$\mathcal{L}(\bm{\theta},\bm{\phi} )=\sum_{i=1}^{8} \|\bm{S}-\bm{\hat{S_i}}\|^2_2$, where $\bm{\hat{S_i}}\triangleq D_{\bm{\phi}}(\bm{\hat{Y}})$ is the intermediate reconstruction of the source signal at the transmitter after each block.  {This additional loss term can guarantee the intermediate transmission quality while incurring a trade-off with the final performance\cite{wu2023transformer}, where the weight assigned to each term can regulate this trade-off.}

The cumulative distribution function of the required bandwidth to achieve two distinct PSNR values ($25$dB and $30$dB) is presented in Fig. \ref{P_ratio}. We can observe that JSCCformer-f provides significant improvements compared to DeepJSCC-f and BPG-Capacity across the entire range, particularly when the target PSNR is higher. Fig. \ref{average_ratio_PSNR} displays the average bandwidth ratio needed to achieve distinct target PSNR values. The observed significant improvements compared to the digital scheme confirm the theoretical results in this practical scenario. Furthermore, the JSCCformer-f model exhibits superior performance and establishes a new state-of-the-art performance, particularly in the high target PSNR value regimes. We can conclude that the JSCCformer-f maintains its superiority in the variable rate transmission scenarios and achieve the state-of-the-art performance.


\subsection{JSCCformer-f for broadcast channels}
\begin{figure}[t]
    \centering
    \centering
    \includegraphics[scale=0.45]{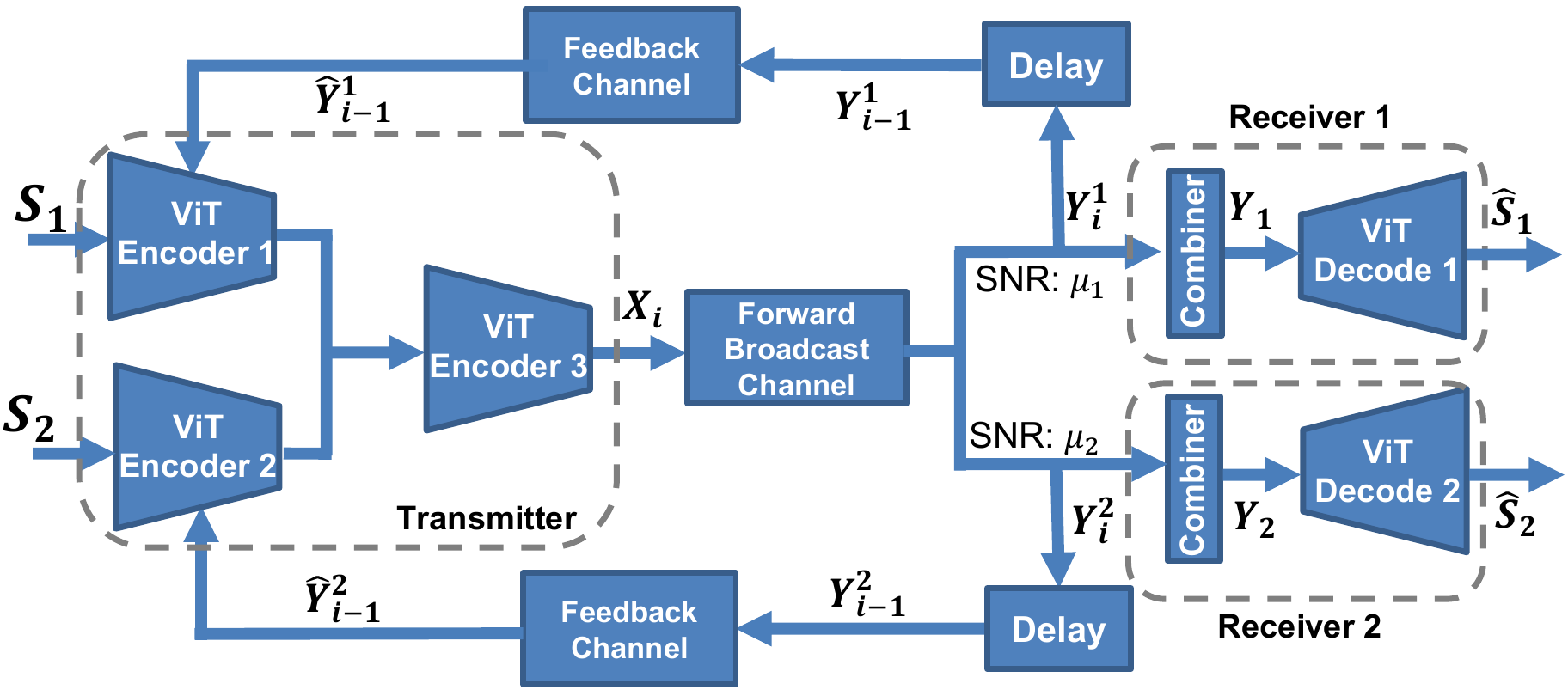}
    \caption{Schematic illustration of extending our JSCCformer-f scheme into broadcast channels.}
    \label{broadcasting_fig}
\end{figure}

This section verifies the generalizability of the proposed JSCCformer-f architecture by extending it to the broadcast channel. In the broadcast channel model, as shown in Fig. \ref{broadcasting_fig}, one transmitter aims to deliver two images $\bm{S_1, S_2}$ to two receivers, respectively. 
The transmitter first encodes the two inputs to a block of symbols $\bm{X_i}$ and then transmits $\bm{X_i}$ through the broadcast channel.
The SNRs associated with the two receivers are $\mu_1$ and $\mu_2$.
We denote the received signal at the two receivers by $\bm{Y^1_i}$ and $\bm{Y^1_2}$, respectively.
Subsequently, $\bm{Y^1_i}$ and $\bm{Y^1_2}$ are fed back to the transmitter for the encoding of the next block, $X_{i+1}$, and so on.
The feedback is assumed to be noiseless in this section.

In contrast to the previous point-to-point scheme, the transmitter in this scheme employs a broadcast code to map the signal source based on the feedback signals from both receivers. To optimize the network efficiency for multi-receiver scenarios, we adopt two compact ViT-encoders $(L_t=4, N_s=4,d=256)$ per message, following Algorithm \ref{encoder_algorithm}, and an additional ViT-encoder $(L_t=4, N_s=4,d=256)$ with the same structure to combine the feature maps of the two source signals. At each receiver, a ViT-decoder $(L_t=4, N_s=4,d=256)$ is deployed to reconstruct the received images $\bm{\hat{S}_i}$ following Algorithm \ref{decoder_algorithm}. The PSNR of receiver $i$ is denoted by $\text{PSNR}_i, i=1,2$. The loss function is given by:
$\mathcal{L}_{2}=\lambda\text{MSE}(\bm{S_1},\bm{\hat{S}_1})+(1-\lambda)\text{MSE}(\bm{S_2},\bm{\hat{S}_2})$, where parameter $\lambda\in [0,1]$ balances the performances of the two receivers.

As the benchmark, we again consider the digital scheme `BPG-Capacity', which utilizes BPG as the source coding scheme and assumes capacity-achieving channel codes of the broadcast channel with feedback \cite{ozarow1984achievable,cover1998comments}. In particular, the achievable rate region lies within the intersection of the curves parameterized by $\alpha\in [0,1]$ as:
\begin{equation}
\begin{array}{c}
     R_1=\frac{1}{2} (1+\frac{\alpha P_s}{\frac{\sigma_1^2\sigma_2^2}{\sigma_1^2+\sigma_2^2}}), 
     R_2=\frac{1}{2} (1+\frac{(1-\alpha) P_s}{\alpha P_s+\sigma_2^2}),
\end{array}
\label{eq_1_broad}
\end{equation}
and 
\begin{equation}
\begin{array}{c}
     R_1=\frac{1}{2} (1+\frac{(1-\alpha) P_s}{\alpha P_s+\sigma_1^2}), 
     R_2=\frac{1}{2} (1+\frac{\alpha P_s}{\frac{\sigma_1^2\sigma_2^2}{\sigma_1^2+\sigma_2^2}}),
\end{array}
\label{eq_2_broad}
\end{equation}
where $R_1$ and $R_2$ are the achievable rates, $\sigma_1^2$, $\sigma_2^2$ are the corresponding noise power of each channel, and $P_s$ is the power constraint.
\begin{figure}[t]
\centering
    \includegraphics[scale=0.42]{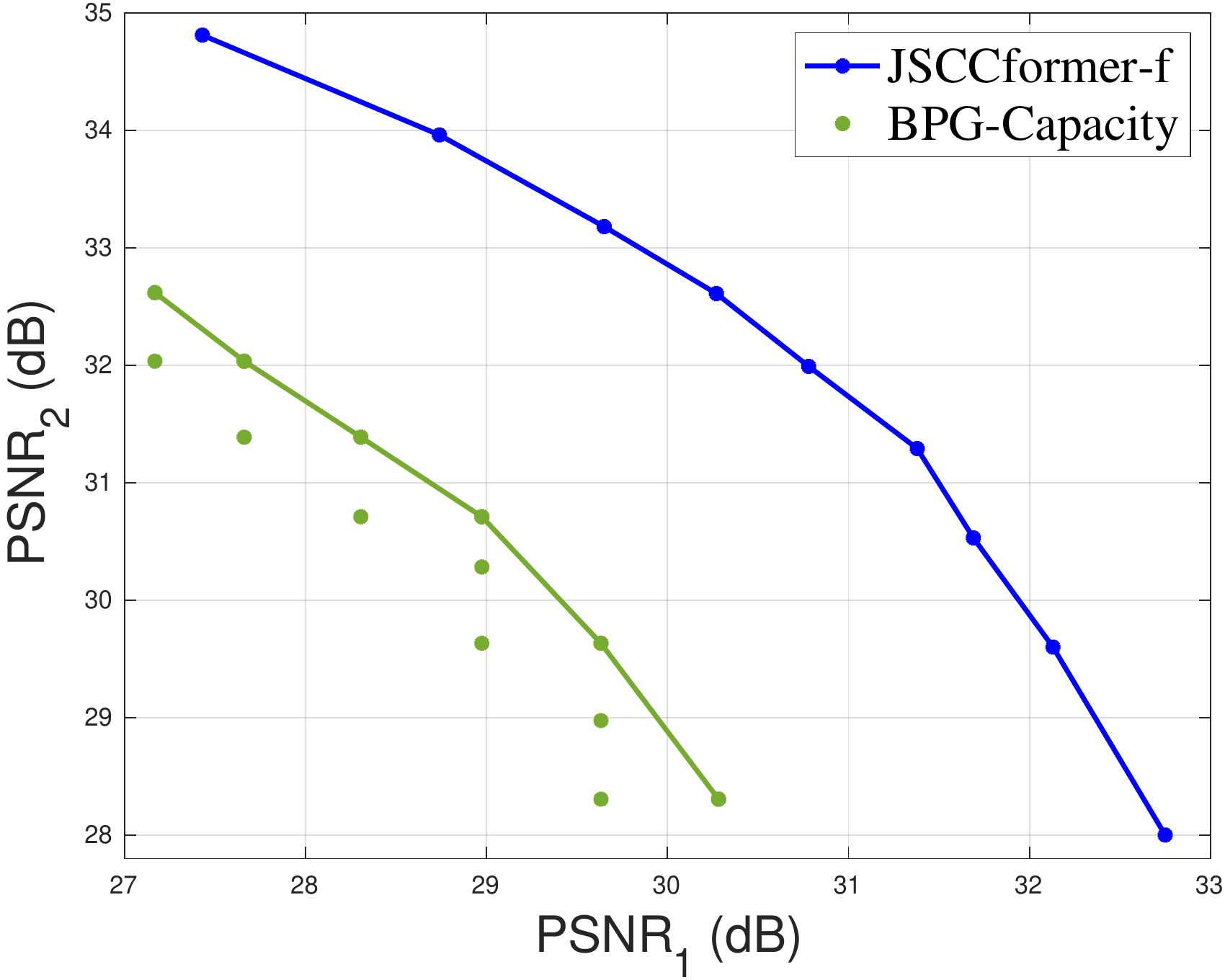}
    \caption{PSNR performance region from different schemes, where $R=1/6$,  $\mu_1=4$ dB and $\mu_2=7$ dB  }
    \label{region_PSNR}
\end{figure}
We first compare the achievable PSNR region of the JSCCformer-f and the BGP-Capacity method in Fig. \ref{region_PSNR}, where we consider $\mu_1=4, \mu_2=7$ (dB) and $R=1/6$. For BPG-Capacity, we plot each PSNR performance pair $(\text{PSNR}_1,\text{PSNR}_2)$, considering the convex hull of the capacity region. For JSCCformer-f, we plot the performance pair from each model trained with different $\lambda\in [0.05,0.95]$. Significant improvements can be observed from our JSCCformer-f-broadcast scheme (at least $3$dB for each receiver).

\begin{figure}[t]
\centering
    \includegraphics[scale=0.45]{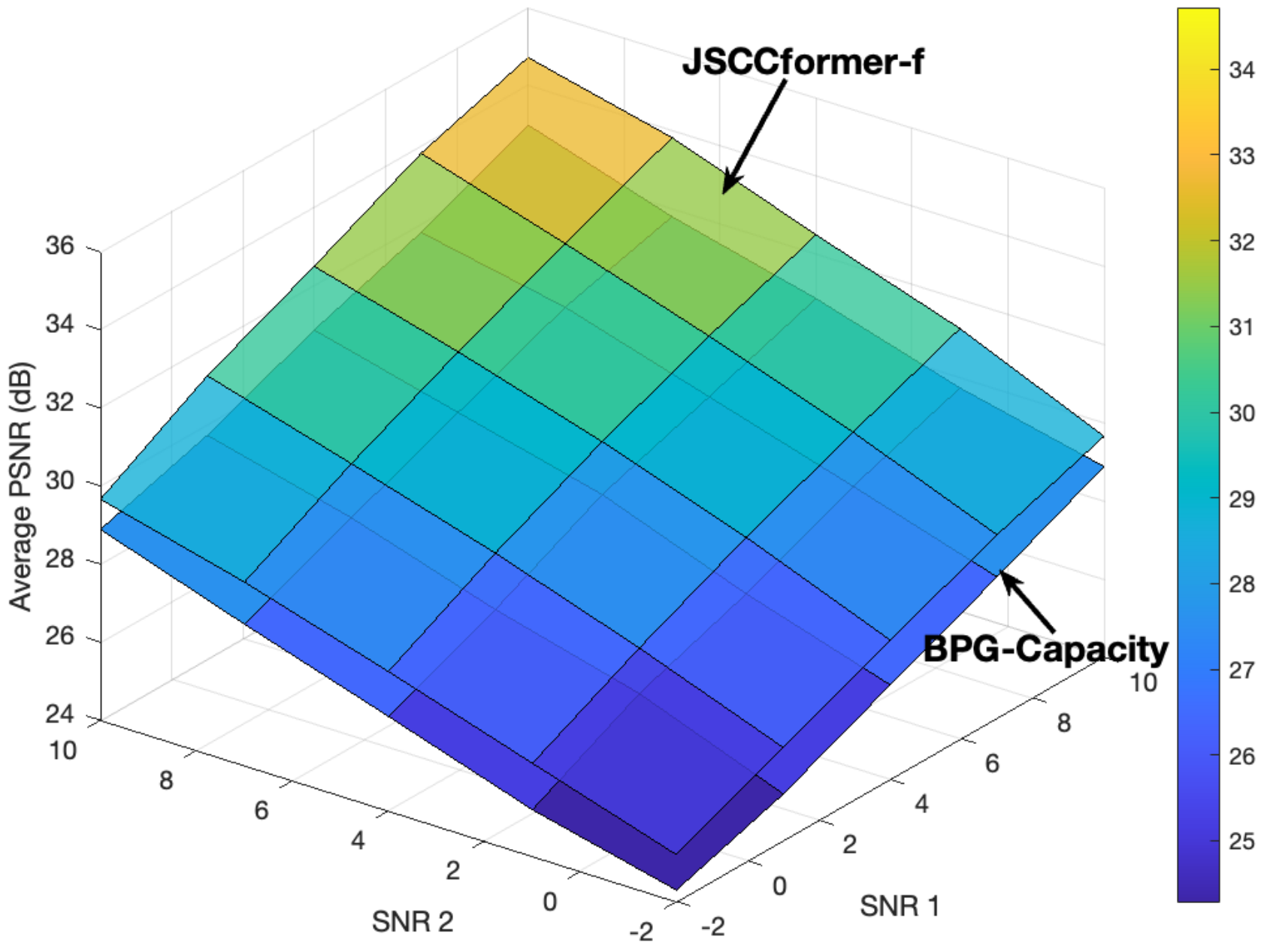}
    \caption{Average PSNR performance of JSCCformer-f (broadcast) for different channel SNR pairs.}
    \label{broadcasting}
\end{figure}

We then consider the average PSNR $(\frac{\text{PSNR}_1+\text{PSNR}_1}{2})$ of JSCCformer-f and BGP-Capacity as a function of various SNR pairs $\mu_1,\mu_2 \in [-2,10]$ dB in Fig. \ref{broadcasting}, where we set $\lambda=0.5$ and $R=1/6$. We can observe that JSCCformer-f outperforms BPG-Capacity in all the SNR pairs. The average PSNR gains are at least $0.9$ dB. In conclusion, the proposed JSCCformer-f framework can be easily generalized to broadcast channels and achieves significantly better results than the conventional separation approach.

\section{Conclusion}
We presented a new feedback-aided wireless image transmission paradigm, dubbed JSCCformer-f.  
JSCCformer-f addresses the four key problems of existing image transmission methods with channel feedback, i.e., high complexity, inadaptability, suboptimality, and non-generalizability.
First, JSCCformer-f generates coded symbols using a unified encoder by exploiting the semantics from the source image and the feedback signal from the receiver (i.e., the decoder's current belief about the source image), and is computationally efficient.
Second, JSCCformer-f shows channel adaptability over a wide range of SNRs without the need for a separate channel estimation module.
Third, JSCCformer-f sets a new state-of-the-art and significantly improves the transmission quality in all the SNRs and bandwidth ratio values considered in this paper, compared with DeepJSCC-f and traditional digital schemes.
Finally, we extended JSCCformer-f to broadcast channels and demonstrated that the transmitter can learn to adaptively map the inputs and the multiple feedback signals to the channel input, considering different channel conditions to satisfy multiple receivers.

Moving forward, JSCCformer-f can be extended in multiple directions:
\begin{itemize}
    \item {We have assumed that the receiver passively feeds back the received coded symbols and relied on the encoder to extract semantic information from the feedback. More generally, the receiver can actively code its feedback based on its knowledge about the source, and can add protection against noise in the feedback channel\cite{ozfatura2022feedback}.}
    \item {The encoder of JSCCformer-f utilizes the SA mechanism to process the received feedback. In principle, more advanced learning techniques, e.g., cross-attention mechanism and contrastive learning methods, can be used to extract useful semantics from the feedback signals and generate coded symbols to refine the receiver's belief.}
    \item {Another interesting line is to consider a generative model available at the receiver\cite{erdemir2022generative}, which may not be known to the transmitter. Feedback in this case needs to convey this additional knowledge to the encoder.}
\end{itemize}



\bibliographystyle{IEEEtran}
\typeout{}
\bibliography{ref}

\newpage
\appendices
\section{Ablation Study}
\subsection{Different positional embedding methods and Siamese layer}
{To provide a more comprehensive analysis of the JSCCformer architecture, we have carried out ablation studies concerning the positional embedding methods and Siamese layer.}

{The results for different positional embedding methods are presented in Table \ref{table_ablation_PE}. We can observe that CPE and DPE exhibit comparable performance for the CIFAR10 dataset. Given the lightweight nature of the DPE block, characterized by its rapid convergence speed with satisfactory performance, we set the DPE as the default positional embedding method for JSCCformer-f. Subsequently, we assess JSCCformer-f on the CelebA dataset. The results presented in Table \ref{table_ablation_PE} show that the model equipped with the CPE block outperforms the DPE-equipped counterpart for high-dimensional image transmission.}

{To better understand the effect of the Siamese layer. We repeat the experiments in Fig. 4a without the Siamese layer. PSNR results presented in Table II clearly show that, similar to \cite{GBAFC,shao2022attentioncode}, the incorporation of a Siamese layer enhances the transmission performance, particularly in low SNR scenarios.
}
\begin{table}[t]
    \caption{{Ablation study over different positional embedding methods.}}
    \centering
    \begin{tabular}{|c|c|c|c|}
            \hline
    \multicolumn{4}{|c|}{CIFAR10 dataset, $R=1/6$ and $m=2$.}\\
            \hline
    {Channel SNR} & $1$dB & $4$dB& $7$dB\\
    \hline
     JSCCformer-f with DPE block & ${28.06}$dB &  \textbf{$\bm{30.64}$dB}& \textbf{$\bm{32.98}$dB} \\

     JSCCformer-f with CPE block & \textbf{$\bm{28.07}$dB} & $30.60$dB& $32.91$dB \\
    \hline
        \multicolumn{4}{|c|}{CelebA dataset, $R=1/12$ and $m=2$.}\\
        \hline
        {Channel SNR} & $1$dB & $4$dB& $7$dB\\
    \hline
     JSCCformer-f with DPE block & $31.81$dB & ${33.34}$dB & ${35.01}$dB \\
     {JSCCformer-f with CPE block} & \textbf{$\bm{32.13}$dB} & \textbf{$\bm{33.86}$dB} & \textbf{$\bm{35.46}$dB}\\
     \hline

    \end{tabular}
    \label{table_ablation_PE}
\end{table}

\begin{table}[h]
    \caption{{Ablation study of the Siamese layer on the CIFAR10 dataset.}}
    \centering
    \begin{tabular}{c|c|c|c}
    {Different models} & $1$dB & $7$dB& $13$dB\\
    \hline
     JSCCformer-f & $\bm{28.06}$dB & $\bm{32.98}$dB & $\bm{36.37}$dB\\
     JSCCformer-f w/o Siamese & $27.45$dB & $32.53$dB & $36.01$dB \\
    \end{tabular}
    \label{table_ablation_siame}
\end{table}

\subsection{Different feedback signals}
Our study assumes that the channel SNR is already known to both the transmitter and receiver, where the transmitter and receiver can select the best models for different channel conditions. If we solely use the estimated SNR as the feedback signal, transmitting the source via multiple blocks with feedback will be meaningless, as the channel condition information from the feedback is already given to both sides.

However, investigating efficient feedback signals in our paradigm is an interesting research direction. We add supplementary ablation experiments over different feedback signals under the same architecture. Specifically, we train the JSCCformer-f with the SNR values as the feedback signal. The performance is presented in Table \ref{table_ablation_snr}.
 
\begin{table}[h]
    \caption{Ablation study for the PSNR performance of JSCCformer-f over different feedback signals, where $R=1/6$, $m=2$}
    \centering
    \begin{tabular}{|c|c|c|c|}
    \hline
    \textbf{Different Feedback Signal} & \textbf{$1$dB} & \textbf{$4$dB}& \textbf{$7$dB}\\
    \hline
     Channel output signals& $\bm{28.06}$dB & $\bm{30.64}$dB & $\bm{32.98}$dB \\
          \hline
     No feedback ($m=1$)& ${27.24}$dB & ${29.70}$dB & ${31.93}$dB \\
     \hline
    SNR with specific training & ${27.16}$dB & ${29.59}$dB & ${31.84}$dB\\
     \hline
     SNR with random training& ${27.09}$dB & ${29.49}$dB & ${31.60}$dB \\
     \hline
    \end{tabular}
    \label{table_ablation_snr}
\end{table}

If JSCCformer-f employs the accurate SNR value as the feedback, the performance is close to the JSCCformer-f model tailed for the same channel condition without the channel feedback. This observation implies that providing solely the accurate average SNR value to the model does not yield performance improvement from the feedback. This is because the average SNR feedback here is just a constant value during different blocks, which cannot convey dynamic information about the current transmission quality to help refine the next block. It tends to ignore this constant SNR value while training the JSCCformer-f for a specific channel condition. Then, the learning process becomes similar to the JSCCformer-f ($m=1$) without feedback scenario. Instead, JSCCformer-f, with the channel output feedback, can dynamically get the decoder's current reconstruction quality and then encode additional channel symbols to refine the performance, thus improving the performance. 
\begin{table}[h]
    \caption{Ablation study with different architecture on Kodak dataset.}
    \centering
    \begin{tabular}{c|c|c|c}
    \textbf{Different JSCC backbones} & $1$dB & $4$dB& $7$dB\\
    \hline
     {JSCCformer-f} & $\bm{30.84}$dB & $32.05$dB & $\bm{33.16}$dB\\
     JSCCformer-f with blocks in\cite{10094735} & $30.69$dB & $\bm{32.19}$dB & $32.84$dB \\
    \end{tabular}
    \label{table_ablation_block}
\end{table}
\begin{table*}[h]
    \caption{{Ablation study over different loss functions for JSCCformer-f on the CIFAR10 dataset, where $m=4$ and $R=1/3$.}}
    \centering
    \begin{tabular}{c|c|c|c|c|c}
    \textbf{Different Channel SNR values } & $1$dB &$4$dB & $7$dB&$10$dB &$13$dB\\
    \hline
      $\|\bm{S}-\bm{\hat{S}_m}\|^2_2$& $\bm{33.11}$\textbf{dB}& $\bm{36.28}$\textbf{dB}& $\bm{38.85}$\textbf{dB}& $\bm{41.17}$\textbf{dB} & $\bm{43.32}$\textbf{dB}\\
      $\sum_{i=1}^{m} \|\bm{S}-\bm{\hat{S}_i}\|^2_2$& ${32.53}$dB & ${35.33}$dB&  ${37.85}$dB&  ${39.84}$dB&  ${41.59}$dB\\
    \end{tabular}
    \label{ablation_over_loss}
\end{table*}
\begin{table*}[h]
    \caption{{Ablation study over different partition methods of JSCCformer-f on the CIFAR10 dataset, where $m=1$ and $R=1/6$.}}
    \centering
    \begin{tabular}{c|c|c|c|c}
    \textbf{Different partition methods } & $1$dB &$4$dB & $7$dB&FLOPs (G)\\
    \hline
      $p=8$, $l=64$, $c=48$& $\bm{27.24}$dB& $\bm{29.70}$dB& $\bm{31.93}$dB&${0.832}$\\
      $p=4$, $l=16$, $c=192$& ${26.12}$dB & ${28.72}$dB&  ${30.80}$dB&${0.210}$\\
      $p=2$, $l=4$, $c=768$& $23.82$dB & ${25.66}$dB&  ${27.35}$dB&${0.054}$\\
      {$p=1$, $l=1$, $c=3072$}& $20.58$dB & ${21.17}$dB&  ${21.84}$dB&${0.015}$\\
    \end{tabular}
    \label{ablation_over_patition}
\end{table*}

\begin{table*}[h]
\caption{Coding time of the tradition models without feedback}
\centering
\begin{tabular}{|c|c|c|c|c|}
\hline
\textbf{Different methods}&{\textbf{\textbf{BPG-LDPC}}} &\textbf{\textbf{JSCCformer-f}}&\textbf{\textbf{JSCCformer-f}} &\textbf{\textbf{JSCCformer-f (lite)}} \\
\textbf{}&{{{(CPU)}}} &{{w/o feedback, (GPU)}}&{{($m=2$, GPU)}} &{{($m=2$, GPU)}} \\

\hline
\textbf{Encoding time} (ms) &{36.34}&{4.75}&{13.76} &{9.04}\\
\hline
\textbf{Decoding time} (ms) &{43.28}& {4.68} & {5.44}&{4.12}\\
    \hline 
\end{tabular}
\label{table_para_tra}
\end{table*}
Interestingly, when we train the model with SNR feedback across the random training SNR values, the model demonstrates channel adaptability to varying channel conditions. This adaptability exhibits the advantage of employing SNR values as channel feedback in our approach. Meanwhile, it proves that ViT can gain the channel-adaptability from the self-attention mechanism.
\subsection{Different backbones}
To validate our model with different ViT-based JSCC backbones, we conduct an ablation study with the backbone from\cite{10094735} over the Kodak dataset with $R=1/12$ and $m=2$. From the Table. \ref{table_ablation_block}, we can observe that the Swin-transformer block can be helpful for better performance when $SNR=4$dB. It shows the potential that some advanced architecture can improve our method in the future.

\begin{figure}[h]
\centering
    \includegraphics[scale=0.4]{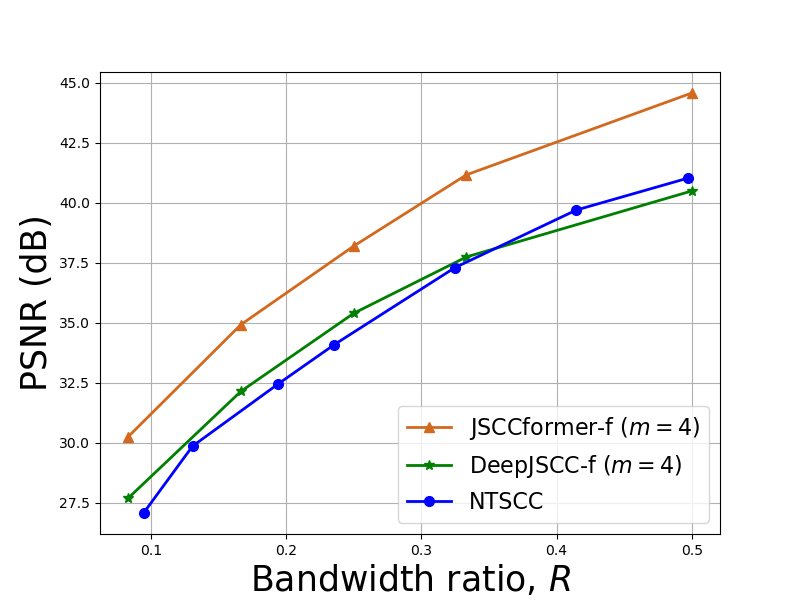}
    \caption{\textcolor{blue}{Performance comparison of different models as a function of the bandwidth ratio $R$ in {AWGN channel} when SNR$=10$ dB with noiseless feedback.}}
    \label{ntscc}
\end{figure}

\textcolor{blue}{We also incorporate comparison with an alternative ViT-based DeepJSCC approach called NTSCC~\cite{9791398} in Figure \ref{ntscc}. We note two primary distinctions of this scheme. Firstly, NTSCC operates without feedback. Secondly, NTSCC employs an adaptive rate transmission strategy, where NTSCC performs coding operations with varying channel bandwidth costs for individual image patches, instead of a fixed bandwidth cost as in \cite{10094735, wu2023transformer}. In some scenarios, this could result in a conservative compression rate, reducing the average bandwidth requirement.} \textcolor{blue}{As depicted in Figure \ref{ntscc}, it is evident that JSCCformer-f consistently outperforms other models across diverse bandwidth ratios.}

\subsection{Addtional loss term}
The introduction of the additional loss term for variable rate transmission essentially imposes increased training demands on the model to accommodate variable rate transmission requirements. The goal is to enhance the quality of each intermediate reconstruction, instead of focusing solely on the final result. In contrast, from an end-to-end perspective, the preceding loss function with a single mean squared error (MSE) term is oriented towards the optimal final reconstruction quality.

To better illustrate the impact of this loss function, we added an ablation study whose results are presented in Table \ref{ablation_over_loss}. Compared with the traditional single loss term, additional intermediate reconstruction constraints generally result in degradations over the final reconstruction performance.

\subsection{Different partition methods}
When there is no feedback, JSCCformer-f ($m=1$) is a pure ViT-based JSCC pipeline. There is a prevalent concept of ``partition operation'' in the recent ViT literature, which is actually a sequentialization operation applied before the encoding stage. We note that, this ``partition operation'' only serves as a preprocessing step and is unrelated to the $m$ blocks of transmission. Different partition methods applied in the ViT-based model, serving as a hyperparameter, can yield different input patch sizes and transmission performance, particularly in scenarios without feedback. Specifically, this partition operation converts the original signal into patch sequences. The following lines from Section III of the manuscript explain the details of this partition operation: ``Specifically, given a source image $\bm{S}\in\mathbb{R}^{h\times w\times 3}$, we divide $\bm{S}$ into a grid of $p\times p$ patches, and flatten the pixel intensities of each patch to form a sequence of vectors of dimension $\mathbb{R}^{\frac{3hw}{p^2}}$. In this way, $\bm{S}$ is converted to $\bm{S_s}\in\mathbb{R}^{l\times c}$, where $l={p^2}$ is the sequence length and $c\triangleq\frac{3hw}{p^2}$ is the dimension of each vector.'' 

To investigate the effect of different partition methods when there is no feedback, we conduct an ablation study whose resultsa are presented in Table \ref{ablation_over_patition}. For an input image with a fixed size, a higher value of $p$ indicates more image patches partitioned from the original image, resulting in longer sequences with smaller dimension elements. This augmentation allows the model to learn the representation over the sequence with fewer dimensions, enhancing the model's capability for encapsulating finer-grained features. However, it comes at the cost of introducing more computation complexity, where the complexity of each attention layer is $O(l^2d)$ with $d$ denoting the intermediate feature dimension.

From Table \ref{ablation_over_patition}, we can suggest $p=8$ as the best suitable partition method for JSCCformer-f over the CIFAR10 dataset, as it provides a very competitive performance with reasonable computation complexity. In general, we observe that an increase in $p$ improves the model's transmission performance at the expense of increased training complexity.

\subsection{Comparison coding time with traditional methods}
There is an inherent trade-off in any feedback channel. Even if the increase in the coding and decoding complexity is ignored, the feedback communication introduces additional delays. Whether exploiting the feedback is worth this additional cost depends on the potential performance gain and application requirements. Additionally, there is also the training and inference (coding) complexity increases. We believe that training complexity is not a big problem since this only needs to be done once, and the obtained code is deployed afterwards. 

The particular structure of our code limits the inference complexity significantly, unlike the DeepJSCC-f code. The parallelizable architecture of our code makes it particularly efficient to be deployed on GPU-like systems. We compare our JSCCformer-f run on the GPU with the traditional separation-based method (BPG-LDPC scheme) run on the CPU. \textcolor{blue}{It is noteworthy that the computational efficiency of GPU implementations should be superior due to its high-speed computing capabilities and parallel processing advantages. However, it is essential to highlight that prevailing conventional transmission approaches mainly rely on CPU implementations, and GPU-based BPG and LDPC methods implementations are not yet accessible.} As shown in Table. X, the BPG-LDPC scheme running on the CPU costs more time than our method optimized in the GPU environment.

\subsection{\textcolor{blue}{Additional experiments for JSCCformer-f (lite)}}
\begin{table}[h]
    \caption{\textcolor{blue}{Additional experiments for JSCCformer-f (lite) for different SNR values and bandwidth ratios, where $m=4$ for $R=1/3$ and $m=2$ for $R=1/6$.}}
    \centering
    \begin{tabular}{p{0.04\textwidth}<{\centering}|p{0.1225\textwidth}<{\centering}|p{0.057\textwidth}<{\centering}|p{0.057\textwidth}<{\centering}|p{0.057\textwidth}<{\centering}}%
    &\textbf{Different methods} & $1$dB & $7$dB&$13$dB\\
    \hline
     \multirow{2}{*}{R=1/6}&JSCCformer-f& \textbf{$\bm{28.06}$dB}& \textbf{$\bm{32.98}$dB} & \textbf{$\bm{36.37}$dB}\\
    & JSCCformer-f (lite)&  ${28.02}$dB & $32.73$dB& ${35.59}$dB \\
    \hline
     \multirow{2}{*}{R=1/3}&JSCCformer-f& \textbf{$\bm{33.11}$dB}& \textbf{$\bm{38.85}$dB} & \textbf{$\bm{43.32}$dB}\\
    & JSCCformer-f (lite)&  ${32.89}$dB & $38.57$dB& ${41.87}$dB \\
    \end{tabular}
    \label{ablation_over_lite}
\end{table}

\textcolor{blue}{To enhance clarity, we conducted additional experiments on JSCCformer-f (lite) across various bandwidth ratios and channel SNR values. The results of these experiments are detailed in Table \ref{ablation_over_lite}. We can observe that the performance of JSCCformer-f (lite) is only slightly worse than the optimal performance of JSCCformer-f at low SNR regimes, making it a promising solution in practice.}

\end{document}